\documentclass[preprint,12pt]{elsarticle}
\usepackage{graphicx}
\usepackage{dcolumn}
\usepackage{bm}

\usepackage{graphics}
 \usepackage{graphicx}
 \usepackage{epsfig}

\usepackage{amssymb}
\usepackage{amsmath}
\usepackage[normalem]{ulem}

\usepackage{color}
\definecolor{darkgreen}{rgb}{0,.6,0}

\journal{Journal of the Mechanics and Physics of Solids}

\begin{document}


\title{Numerical investigation of the interaction between the martensitic transformation front and the plastic strain in austenite}

\author{Julia Kundin\corref{cor1}\fnref{label1}}
\ead{Julia.Kundin@uni-bayreuth.de}
\author{Evgeny Pogorelov\fnref{label2}}
\author{Heike Emmerich\fnref{label1}}
\address[label1]{Material and Process Simulation (MPS), University Bayreuth, 95448 Bayreuth, Germany}
\address[label2]{Advanced Ceramics, University Bremen, 28359 Bremen  Germany\fnref{label2}}
\cortext[cor1]{Corresponding author}

\date{\today}

\begin{abstract}
Phase-field simulations of the martensitic transformation (MT) in the austenitic matrix, which has already undergone the plastic deformation, are carried out. 
For this purpose the elasto-plastic phase-field approach of incoherent MT developed in the previous work [Kundin et. al. J. Mech. and Phys. Solids 59 (2011) 2012] is used. 
The evolution equation for the dislocation density field is extended by taking into account the thermal and athermal annihilation of the dislocations in the austenitic 
matrix and the athermal annihilation at the transformation front. It is shown that the 
plastic deformation in the austenite caused by the MT interacts with the dislocation field and the MT front that leads to the inhomogeneous increasing of the total dislocation 
density. During the phase transformation one part of the dislocations in the austenite is inherited by the martensitic phase and this inheritance depends on the kinetics and the crystallography of MT.  Another part of dislocations annihilates at the transformation front and decreases the dislocation density in the growing martensite.  
Based on the simulation results the specific type of phenomenological dependency between the inherited dislocations, the martensite phase fraction and the plastic deformation is proposed.
 \end{abstract}

\maketitle

\section{Introduction} \label{section_Introduction}

The morphology of the martensitic microstructures is strongly related to the mechanical properties of steel and alloys. It is important to predict the formation of the
martensitic microstructure precisely. Furthermore, during the MT the incremental elastic strain energy is reduced by the
formation of a heterogeneous array of different orientation variants of the martensitic phase and by plastic accommodation. The development of 
appropriate models allows to calculate the strains associated with the phase transformation. The understanding and control of localized plasticity and its interaction
with the transformation front is quite essential e.g. for TRIP-steels and shape memory alloys \cite{Dadda2008,Graessel2000}.

Recently, phase-field (PF) models have been extensively studied as a powerful tool for predicting 
microstructural evolution and applied to the martensitic transformation.
Khachaturyan and co-workers developed the phase-field microelasticity (PFM) theory  \cite{Wang1997,Jin2001}, 
which integrates the microelasticity into the phase-field model employing the
fast Fourier transform algorithm. The model was further applied to investigate the
MT in single crystals \cite{Artemev2001} and polycrystalline systems \cite{Artemev2002} 
as well as in multilayer systems under applied stresses \cite{Artemev2000}. 
It has been successfully applied to various coherent phase transformations 
including the prediction of many complicated strain-induced 
morphological patterns \cite{Wang2000,Chen2002,Shen2005,Zhang2007}.  
  Resolving the individual martensite plates the PF approach differs from the phenomenological
martensitic transformation modeling of irreversible processes in various representations:  local, crystalline and mean-field \cite{Kubler11,Ostwald2011,Ostwald2012,Fischer98,Fischer2000,Idesman2000,Kaganova89}, 
which are widely used in the simulation of shape memory alloys and TRIP-steels due to their simplicity and efficient incorporation of elasto-plastic effects on mesoscale. To understand the mechanism of nucleation and growth conditions of martensitic plates inside an inhomogeneous plastic strain field the number of theoretical models were developed \cite{Olson86,Cherkaoui2000}. These models include the dislocation theory of the martensitic interfacial structure and introduce a driving force of interface propagation in inelastic materials.

Besides the coherent phase transformations the MT in technical alloys is usually associated with plastic strains. 
The effect of plasticity, which can be represented as the generation and the motion of dislocations during the MT, is very complex. On the one hand, if the MT is caused 
by applied loads, the evolution of the plastic deformation reduces the driving force of the transformation. On the other hand, a local plastic relaxation allows the 
accommodation of the strain caused by the MT and the nucleation and growth of new martensitic variants. Finally, statistically stored dislocations are an irreversible 
structural change which affect the energy landscape by their own elastic fields.     
Thus, a general phase-field model for MT should include not only the driving forces originating from the elastic fields, but also the effects resulting from the plastic 
strain fields.

The plastic activity in the phase-field theory has already been treated by modeling the individual dislocations \cite{Hu2001} 
and their  coupling dynamics \cite{Rodney2001,Rodney2003,Wang2001,Zhou2007,Koslowski}. A  phase-field model of the evolution of a dislocation system due to the evolution of 
dislocation order parameters based on the time depending Ginzburg-Landau (TDGL) equation was developed by Wang et~al. \cite{Wang2001}. 
At the same time, Koslowski et~al. \cite{Koslowski} 
formulated a phase-field theory of the dislocation dynamics for an arbitrary number and arrangement of  dislocation lines based on a general framework for dissipative systems. 

 The plastic deformation can also be added to the phase-field model by introducing a 
plastic strain field defined at the mesoscale. A version of this approach has been recently proposed by Zhou et al. \cite{Zhou2008,Zhou2010}, 
where the plastic strain is related to the inter-dislocation distance. The evolution of the dislocation phase fields is described by the TDGL equation similar to the evolution of the martensitic phase fields, where the driving force is the elastic shear energy density relaxed by the plastic strain. 
The difference to the individual dislocation dynamic models is that the model proposed by Zhou et al. describes the evolution of dislocation order parameters related to the slip systems of the crystal. 

 In recent years a similar approach for the simulations of the evolution of the martensitic phase transformations and the dislocation order parameters by the 
TDGL equations was developed by Levitas and Javanbakht \cite{Levitas12}. The phase field approach was applied to interaction
of the phase transformation and the individual dislocation evolution \cite{Levitas13c}.
 As a result  a number of model problems of the stress-induced phase-transformations interacting with dislocation evolution were solved.
 It is also important that in the work \cite{Levitas13b} the general phase-field theory for multivariant martensitic phase transformations and explicit models have been
formulated for the most general case of large strains. The general thermodynamic approach allows to determine the driving force for the change of the order parameters and 
the boundary conditions for the order parameters. Moreover, it allows to calculate correct interface stresses and to elucidate the importance of the interface width  
\cite{Levitas13a}. Remarkable is that Levitas et~al. previously already developed a finite-strain-continuum thermomechanical approach to simulate the transformation-induced stress and plastic strain fields \cite{Levitas02} based on a theory of martensitic phase transformations developed earlier by Levitas \cite{Levitas2000}.

 Another simplified version of this approach has been derived by Yamanaka  \cite{Yamanaka07,Yamanaka2009} based on the PF microelasticity theory and the elasto-plastic 
 PF model suggested by Guo et al. \cite{Guo2005} which not take into account different slip systems. 
 In recent years elasto-plastic phase-field models of the martensitic transformation using similar evolution equations for the plastic strain have been developed by Koyama's group \cite{Cong12a,Cong12b,Cong13}.  This approach also follow H.K. Yeddu et~al. \cite{Yeddu12a,Yeddu12b,Yeddu13,Malik12} 
 for the phase-field modeling of the martensitic microstructure evolution in steels by using the elasto-plastic finite-element method.  Further, Yamanaka et~al. \cite{Yamanaka12} 
 proposed  the model to describe the austenite-to-ferrite transformation, where the mesoscale crystal plasticity theory is used \cite{Pan1983}. They combine the crystal plasticity
finite element method with the multi-phase-field method to simulate the austenite-to-ferrite transformation with  the diffusion of carbon atoms
in low carbon steels.

To describe the plastic deformation in single crystals many models use the crystal plasticity framework in which the evolution of plastic strain is described by means of 
an elastic driving force.   A microstructural strain-hardening model suitable for
crystal plasticity simulations developed by Ma and Roters \cite{Roters2000,Roters2010,Ma2004}  has recently made remarkable progress and allows the numerical study of deformation processes on 
the basis of the thermally activated dislocation evolution. The model preserves the crystallographic features of dislocation slip processes and captures the commonly 
accepted concepts of dislocation processes in the plastic deformation, especially various dislocation interaction processes as interactions of mobile and statistically 
stored dislocations, the formation of locks and dipoles or the thermal and athermal annihilation in a continuum dislocation density framework. The original concept 
\cite{Ma2004} for fcc single crystals has been extended to polycrystals, considering grain boundary interactions and geometrically necessary dislocations as well as 
the extension to dislocations in  bcc crystals \cite{Ma2006a,Ma2006b,Ma2007,Roters2010}. It should be noted that a number of other works have been published 
recently, which couple the phase-field simulation with crystal plasticity schemes at the mesoscale  \cite{Takaki2007,Gaubert2010,Takaki2010,Yamanaka13} and do not 
take into account the plastic accommodation caused directly by the solid-solid transformation. 

Crystal plasticity models, their further developments based on microscopically interpretable state-variables and their evolution based on sound physical principles are 
very attractive for the coupling with solid-solid phase transitions and offer new research directions and new insights into the mechanisms of martensitic phase 
transformations. One contribution in this direction is the work of Kundin~et al. ~\cite{Kundin11b}, where the crystal plasticity model 
is coupled to the phase-field model of the MT that allows to resolve the dynamics of individual martensitic plates.   The coupling model also uses methods 
of earlier models for the simulation of the transformation induced plasticity \cite{Tjahjanto2008}.  
 It has been demonstrated that the formation of special martensite morphologies (butterfly type) in a Fe-31$\%$Ni is assisted by the formation of plastic strain fields in the 
 austenitic matrix, which then are inherited by the growing martensitic plates.  This phenomenon was experimentally observed in the works \cite{Salje2009,Sato2009}  and investigated numerically later in  work \cite{Levitas12}.

The main goal of the current paper is to understand the evolution of the dislocations during the MT and the interaction of the plastic strain with the transformation front. 
The kinetic of the MT is described by the elasto-plastic PF model of the MT.  The dislocations are treated as separate dislocation fields. We consider their evolution using 
the model of thermally activated dislocation motion \cite{Roters2000,Roters2010,Ma2004}.  Furthermore,  we take into account three main processes: the generation and annihilation of the dislocations in the austenitic matrix and the annihilation of the dislocations on the propagating martensitic front.   In our investigation we concentrate on the inheritance of the dislocations by the martensitic phase.  The previous work in this direction is the paper of Ostwald et al.~\cite{Ostwald2011}, where the authors introduced an inheritance probability function for the quantitative  description of the inheritance process, which is equal to the fraction of the inherited dislocation density. To describe the inheritance mechanism in our model the probability of the dislocation inheritance is defined as the fraction of the remaining dislocation density after the annihilation on the transformation front. 

We first start with recalling the phase-field microelasticity model of the MT extended by taking into account plastic strains in Section 2. The differential equation describing the evolution of plastic strains and dislocation density is presented in Sections 2.3. The coupling of the 
phase-transformation and the plasticity effects is incorporated in one algorithm. 
Details on the numerical investigation of the interaction between the transformation front and the plastic strain are provided in Section 3, where the model is applied to a a Fe-30$\%$Ni alloy. In addition the resulting effect on the inheritance kinetics was predicted by the suggested model 
and compared with the reported data, when various phase-field model parameters are applied.

\section { Theoretical part} 

\subsection{The main parameters of the elasto-plastic phase-field model of MT}

Following the concept developed in the PFM theory \cite{Wang1997,Artemev2001,Jin2001,Zhang2007} we consider a coherent multi-phase
mixture with the order parameters $\eta_p(\mathbf{r})$ and  the local stress-free strain tensor $\varepsilon^{0}_{ij}(p,
\mathbf{r})=\varepsilon_{ij}^{0}(p)\eta_p(\mathbf{r})$ where $p$ identifies a crystallographic variant. The order parameters are treated as phase-field 
variables so that $\eta_p\in [0,1]$ and satisfy the property $\sum_{p=1}^\nu \eta_p(\mathbf{r})=1-\phi_A(\mathbf{r})$ where $\phi_A(\mathbf{r})$ is the phase-field variable responsible for the presence of the austenite. 

The dislocations are characterized by a Burgers vector $\mathbf{b}^{(\alpha)}$ and a slip plane with a normal $\mathbf{n}^{(\alpha)}$  defined by the crystallography of the austenitic (or martensitic) phase. Here, $\alpha$ is an index corresponding to a slip system.
We can define a dimensionless dislocation function  through the dislocation densities of mobile, $\rho_{M,\alpha}$, and immobile, $\rho_{I,\alpha}$,  dislocations 
\begin{equation}
\phi_{\alpha}^d(\mathbf{r})= b^{(\alpha)}\sqrt{\rho_{I,\alpha}(\mathbf{r})+\rho_{M,\alpha}(\mathbf{r})}\label{DislFunc}
\end{equation}
 as  an analog of a phase-field variable. It is equal to the unity if the distance between the dislocations in the full volume is equal to the Burgers vector.

The local stress-free plastic strain tensor
caused by moving dislocations of an $\alpha$-th slip system 
is given then by

\begin{eqnarray} \label{DislDistributionSimple}
 \varepsilon_{ij}^p (\alpha,\mathbf{r}) 
  &=&   \mathrm{M}_{ij}(\alpha) \gamma_{\alpha}(\mathbf{r}).
\end{eqnarray}
where $\gamma_{\alpha}$ is the plastic shear and $\mathrm{M}_{ij}(\alpha)$ is the Schmid tensor for a slip system $\alpha$ defined as $\mathbf{\hat{M}}^{(\alpha)}=\mathbf{n}^{(\alpha)}\otimes\mathbf{m}^{(\alpha)}$
where $\mathbf{m}^{(\alpha)} = \mathbf{b}^{(\alpha)}/b^{(\alpha)}$ is the unit vector in the direction of the
Burgers vector, which expresses the slip direction.

\subsection{The evolution of the order parameters in the elasto-plastic phase-field model}

The formulation of the total free energy of a system as a function of the order parameters is the key step in the development of phase-field models. In the standard phase-field formulation the total free energy is divided into three terms: the gradient energy term, the double well potential term (both terms are responsible for the interface energy) and the chemical free energy that provides the chemical driving force for the transformation. Concerning the MT, there are an additional elastic strain energy that suppresses or accelerates the transformation.
The chemical free energy and the double well potential term in PFM theory of the MT are usually considered together and  approximated by a Landau polynomial expansion with respect to the order parameters \cite{Toledano1996}. 

In the present study we reformulate the PFM theory in terms of the standard phase-field modeling \cite{Karma98}.  
The evolution of the order parameters in the system with the plastic strain is governed by the elasto-plastic phase-field kinetic equation

\begin{eqnarray}
\frac{1}{\mathcal{M}}\frac{\partial\eta_{p}(\mathbf{r})}{\partial t}&=& K\nabla^2\eta_{p}(\mathbf{r})-H f'_{\eta_{p}}+ \Delta f g'_{\eta_{p}} +\sigma_{ij}^{\text{appl}}\varepsilon_{ij}^0(p)g'_{\eta_{p}} -\sum_{\beta=1}^{\mu}\omega (\phi_{\beta}^d(\mathbf{r}))^2 g'_{\eta_{p}}\nonumber\\
& -&c_{ijkl}\varepsilon_{ij}^0(p)\biggl[\sum_{q=1}^{\nu}\varepsilon_{kl}^0(q)\left(\eta_{q}(\mathbf{r})-\bar{\eta}_{q}\right) + \sum_{\beta=1}^{\mu}\mathrm{M}_{kl}(\beta)\left(\gamma_{\beta} (\mathbf{r}) -\bar{\gamma}_{\beta}\right)\biggr]\nonumber\\
& +&\int c_{ijkl}\varepsilon_{kl}^0(p)e_{i}\Omega_{jm}(\mathbf{e})\hat{\sigma}_{mn}^{0}(\mathbf{k}) e_n \text{e}^{i\mathbf{k}\cdot\mathbf{r}}\frac{d^{3}\mathbf{k}}{(2\pi)^{3}}, \label{KineticEq}
\end{eqnarray}

where the Fourier transform of the stress is calculated by
\begin{equation} \label{F_stress}
\hat{\sigma}_{mn}^{0}(\mathbf{k})=\int_{V} c_{mnjk}\biggl[\sum_{q=1}^{\nu}\varepsilon_{jk}^0(q)\left(\eta_{q}(\mathbf{r}) - \bar{\eta}_{q}\right) +\sum_{\beta=1}^{\mu}M_{jk}(\beta)\left(\gamma_{\beta}(\mathbf{r}) -\bar{\gamma}_{\beta}\right)\biggr]\text{e}^{-i\mathbf{k}\cdot\mathbf{r}}\, d^3\mathbf{r}.
\end{equation}
Here $c_{ijkl}$ are the components of the elastic modulus tensor, $\Omega_{jm}$ is the Fourier transform of the Green function, $\bar{\eta}$ and $\bar{\gamma}$ are the volume averaging functions, $\sigma_{ij}^{\text{appl}}$ is the applied stress,   $\nu$ is the number of the martensitic variants and $\mu$
is the number of the slip systems. The term $\omega (\phi_{\beta}^d(\mathbf{r}))^2$ is responsible for the resistance exerting by dislocations, $\omega$ is a scaling parameter.

The first term on the right hand site of eq.~(\ref{KineticEq}) is the gradient term, which forces interfaces to have a
finite width. The second term is the double potential function and the third term is the driving force. Then $\mathcal{M}$ is the mobility of the interface, $K=\frac{\sigma \xi}{a_1}$ is the gradient coefficient, $H=\frac{\sigma}{a_1\xi}$ is the coefficient before the double well function, $\xi$ and $\sigma$ are the interface width and the interface energy, respectively, which are assumed to be equal for all solid-solid interfaces.

The functions $f'_{\eta_{p}}$ and $g'_{\eta_{p}}$ are the derivatives of the model functions with respect to the chosen order parameter. The model function $f$ is the double well potential and is defined as follows
\begin{eqnarray}
f(\vec{\eta})=\sum_{q=1}^\nu\eta_{q}^2- 2\sum_{q=1}^\nu\eta_{q}^3 +\left(\sum_{q=1}^\nu\eta_{q}^2\right)^2.
\end{eqnarray}
The function $g$ is responsible for the chemical driving force and defined as
\begin{eqnarray}
g(\vec{\eta})=4\sum_{q=1}^\nu\eta_{q}^3-3\left(\sum_{q=1}^\nu\eta_{q}^2\right)^2.
\end{eqnarray}
These model functions serve the goal to keep the order parameters between 0 and 1. 
In Appendix A we show that the chosen model functions satisfy the properties of the standard phase-field model and derive their
relation to the Landau polynomial expansion usually used in PFM theory.

The undercooling responsible for the chemical driving force is given by
\begin{equation}
\Delta f=Q_M(T_M-T)/T_M,
\end{equation}
where $Q_M$ and $T_M$ are is the latent heat  and $T$ is the temperature of the MT, respectively.

\subsection{The evolution of the dislocation density and the plastic strain}
\label{FormationDislocation}

In this section we describe the time evolution of the dislocation density generated
 during the MT and the evolution of the plastic deformation strain. The plastic strain and the dislocation density will first evolve in the austenitic phase. Then, due to the interaction with the transformation front, one part of the dislocations will cross the interface without or 
with small changes of the slip system and will be inherited in the martensite. The other part of the dislocation will annihilate by the interaction with the dislocations of the same slip system collected at the transformation front. The kinetics of these processes depends on the growth velocity of the transformation front and the resistance stress of the matrix.  It should also be taken into account that a part of the dislocations annihilates in the bulk phase. In the following we consider a single crystal system, i.e. a system which contains only one grain without boundaries.

 To describe the evolution of the dislocation density in time and space a reaction-diffusion model was developed by Walgraef and Aifantis \cite{Walgraef85}. They proposed a general balance equation for dislocation dynamics
\begin{equation}  \label{Balance}
\dot{\rho}_{I} + \text{div } \bf{j}_\rho=\text{reactions},
\end{equation}
where $\bf{j}_\rho$ is the dislocation flux.
  In past years this model has initiated the development of many analogous models.  The limitations of these models were considered in \cite{Kubin03} and it was shown that if the homogenization length for the dislocation density is taken smaller than the critical reaction distance between the dislocations the transport term
at the left-hand side of Eq. (\ref{Balance}) becomes very important. In our simulations in section \ref{Section_Numerical_simulation} we have a system size in the order of 0.1 $\mu$m and the distance between the dislocations in the beginning of the microstructure evolution is larger or in the same order. During the MT the distance decreases down to 10 nm, that is in the order of martensitic layer size. But it is still larger than the interface width ($1.5\,l_0$ = 2 nm) and discrete dislocations should be used. This scaling is typical for all previous elasto-plastic phase-field 
models using the continuum dislocation fields.  In order to minimize this drawback we suggest to take into account additional gradient terms in  the evolution equation (see below).

In this section we use the theory of the dislocation evolution presented in \cite{Roters2000,Ma2004} and extend it by inserting dislocation flux terms in the gradient form.
 we define additionally to the mobile and immobile dislocations the parallel dislocation density $\rho_{\mathrm{P,\alpha}}$ and the forest dislocation density $\rho_{\mathrm{F,\alpha}}$ for a slip system $\alpha$ as
\begin{equation}  \label{FDislocation1}
\rho_{\mathrm{P,\alpha}}=\sum_{\beta=1}^\mu\rho_{\mathrm{I,\beta}}\left|\sin\left(\mathbf{n}^{(\alpha)},\mathbf{n}^{(\beta)}\times\mathbf{m}^{(\beta)}\right)\right|,
\end{equation}

\begin{equation}  \label{FDislocation0}
\rho_{\mathrm{F,\alpha}}=\sum_{\beta=1}^\mu\rho_{\mathrm{I,\beta}}\left|\cos\left(\mathbf{n}^{(\alpha)},\mathbf{n}^{(\beta)}\times\mathbf{m}^{(\beta)}\right)\right|.
\end{equation}

The density of the mobile dislocations is calculated as 
\begin{equation}  \label{FDislocation2}
\rho_{\mathrm{M,\alpha}}=\frac{2k_{\mathrm{B}}T}{c_1c_2c_3Gb^3_{\alpha}}\sqrt{\rho_{\mathrm{F,\alpha}}\rho_{\mathrm{P,\alpha}}}, 
\end{equation}
where $G$ is the shear modulus, $k_{\mathrm{B}}$ is the Boltzmann constant, $T$ is the temperature and $c_1$,$c_2$,$c_3$ are material constants. The density of the mobile dislocations collected on the transformation front, $\rho_{M}^{f}$,  is  calculated as an average dislocation density in the austenite over space and time.

Then we calculate the evolution of the immobile dislocation density as
\begin{eqnarray}\label{DensEvolution}
\dot{\rho}_{I,\alpha}&=&c_{4}\sqrt{\rho_{F,\alpha}}\dot{\gamma}_{\alpha}-c_5\rho_{I,\alpha}\dot{\gamma}_{\alpha} -c_7\exp\left(-\frac{Q_{\text{bulk}}}{k_BT}\right)\frac{|\tau_\alpha|}{k_BT}(\rho_{I,\alpha})^2\dot{\gamma}_{\alpha}^{c_8}\nonumber\\
&-&c_9\rho_{I,\alpha}\rho_{M}^{f} \sum_{p=1}^\mu\dot{\eta}_{p} + c_{10}\phi_A \Delta \rho_{I,\alpha},
\end{eqnarray}
where $\dot{\gamma}_{\alpha}$ is the plastic shear rate, $Q_{\text{bulk}}$ is the activation energy for climb, $\tau_\alpha$ is the external stress.

Eq.~(\ref{DensEvolution}) combines three processes in the bulk phase: immobilization of the mobile dislocations, non-thermal annihilation with the immobile dislocations of the same slip system and thermal annihilation by climb of edge dislocations  with material constants $c_{4}$, $c_5$ and $c_{7}$, respectively. The probability of these processes are governed by the plastic shear rate $\dot{\gamma}_{\alpha}$.

\begin{figure}
\begin{centering}
\includegraphics[scale=0.45]{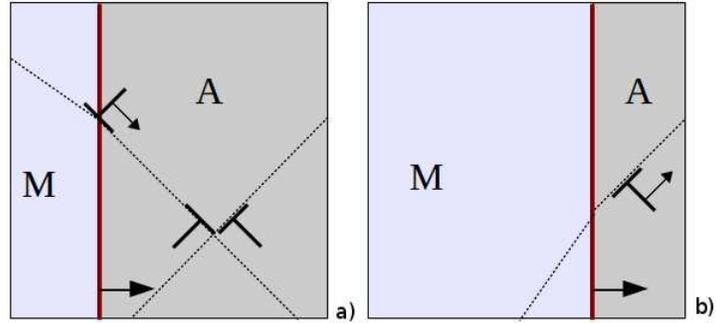}  
\caption{\label{Fig1} 
Schematic illustration of the annihilation process on the transformation front between martensite (M) and austenite (A); (a) before the annihilation, (b) after the annihilation. }
\end{centering}
\end{figure}

  Furthermore, we propose an additional term with a constant $c_9$, which is responsible for the non-thermal annihilation of the mobile dislocations at the martensitic transformation front with  immobile dislocations in the matrix. The probability of this process is proportional to the local dislocation densities and the velocity of the MT front, which is related to the evolution of the order parameter $\dot{\eta}_p$.  The physical meaning of this term is that the transformation front collects the mobile dislocations, which can later annihilate with the immobile dislocations of the same slip system during the front propagation. The released mobile dislocations will be pushed forward by the MT front. This process is illustrated in Fig.~\ref{Fig1}. The collection of the dislocations on the front should rise with the increasing transformation rate. The parameter $c_9$ is a phenomenological parameter, which should be justified  experimentally. In the following we will call the annihilation at the MT front as the 
annihilation term. 
  
 Note that the model parameter $c_9$ is strongly connected to the interface width $\xi$ due to the integration over the phase-field diffuse interface. That is why an appropriate choice of the interface width $\xi$ which is related to the discretization size $\Delta x$ is very important.
  The diffuse interface in phase-field model is used for the modeling of the moving boundary. We use the properties of the diffuse interface to simulate the annihilation process on the moving transformation front. To estimate the average annihilation rate we should take the integral of the annihilation term over the simulation domain. The average annihilation term will depend on the product $\xi \bar{\dot{\eta}}_p$ which is the average front velocity and will be inverse proportional to the distance between the martensite-austenite boundaries, $1/L$.  Hence the annihilation term depends on physical parameters of the system and does not depend on numerical effects.

The last gradient term in (\ref{DensEvolution}) is added as the transport term required in the reaction-diffusion model \cite{Walgraef85} to guarantee the homogeneous distribution of the dislocations and a smooth transition of the dislocation field
profile across the martensite-austenite boundaries. A similar term was used in the models which deal with the dislocation evolution by TDGL equation  \cite{Zhou2010,Cong12b,Cong13}. 

The plastic shear rate can be defined according to the Orowan equation as
\begin{equation}\label{SlipEvolution}
  \dot{\gamma}_{\alpha}=\phi_{A}\rho_{M,\alpha}b_{\alpha}v_{\alpha} ,
\end{equation}
where we assume that the plastic deformation is hindered in the martensite due to much higher flow stresses compared to the austenite. We describe the evolution of plastic slip during the process only in the austenitic phase and  multiply the plastic shear rate with $\phi_{A}$.

The average dislocation velocity $v_{\alpha}$  based on the model of thermally 
activated dislocation motion  can be found as \cite{Ma2004,Roters2000,Roters2010} 
\begin{eqnarray} \label{SlipVelocity}
 v_{\alpha}=\lambda_{\alpha}\nu_{\alpha}\exp\left(\frac{|\tau^{\alpha}|-\tau_{\mathrm{pass}}^\alpha}{k_{\mathrm{B}}T}V_{\alpha}\right),& & \text{    for  } \lvert\tau^{\alpha}\rvert>\tau_{\mathrm{pass}}^\alpha\\\nonumber
& &v_{\alpha}=0, \text{  otherwise}
\end{eqnarray}
where $ \tau_{\mathrm{pass}}^\alpha=c_{1}Gb_{\alpha}\sqrt{\rho_{\mathrm{P,\alpha}}+\rho_{\mathrm{M,\alpha}}}$
is the passing stress of the mobile dislocations, $\nu_{\alpha}=\nu_0\exp\left(-\frac{Q_{\mathrm{slip}}^\alpha}{k_{\mathrm{B}}T}\right)$, $\nu_0$ is the attack frequency,  $Q_{\mathrm{slip}}^\alpha=Gb_{\alpha}^3/2$  is the effective activation energy for the dislocation slip,  $\lambda_{\alpha}=\frac{c_2}{\sqrt{\rho_{F,\alpha}}}$ is the jump width and  $V_{\alpha}=c_3 b_{\alpha}^2\lambda_{\alpha}$ is the activation volume,
where $c_1$, $c_2$ and $c_3$ are material constants.

In the simulations numerical problems arise with eq.~(\ref{SlipVelocity}) due to the very strong exponential function of the stress. For numerical reasons we prefer an alternative variant of the evolution equation according to a Norton type flow rule
\begin{eqnarray}\label{SlipEvolution2}
 v_{\alpha}=\lambda_\alpha \nu_{\alpha}\left(\frac{|\tau^{\alpha}|-\tau_{\mathrm{pass}}^\alpha}{\tau_{\mathrm{cut}}^{\alpha}}\right)^n,& & \text{    for  } \lvert\tau^{\alpha}\rvert>\tau_{\mathrm{pass}}^\alpha\\\nonumber
& &v_{\alpha}=0, \text{  otherwise}
\end{eqnarray}
where $\tau_{\mathrm{cut}}^{\alpha}=1.0$ MPa was calculated by $\tau_{\text{cut}}^{\alpha}=k(\rho_M^{\alpha}\lambda_\alpha\nu_{\alpha} b_{\alpha})^{1/n}$ using the Norton law coefficient  $k=0.2$ GPa s$^{1/n}$ and $n=5$ for steels \cite{Ambroziak2007}.
For the calculation we taken the maximal dislocation density $\rho_F=\rho_P=10^{15}$ 1/m$^{2}$ and the parameters in Table \ref{Table1}.

The external stress $\tau^{\alpha}$, which is the driving force of the dislocation evolution, is caused by the applied force,  growing martensitic phases and dislocation fields. It can be derived from the total 
elastic energy   as 

\begin{eqnarray}\label{TauExternal}
\tau_{\alpha}&=&-c_{ijkl} M_{ij}(\alpha)\biggl[\sum_{q=1}^{\nu}\varepsilon_{kl}^0(q)\left(\eta_{q} - \bar{\eta}_{q}\right)+\sum_{\beta=1}^{\mu}M_{kl}(\beta)\left(\gamma_{\beta}-\bar{\gamma}_{\beta}\right)\biggr]\nonumber\\
&+&\int_{\hat{V}} c_{ijkl}M_{kl}(\alpha)e_{i}\Omega_{jm}(\mathbf{e})\hat{\sigma}_{mn}^{0}(\mathbf{k})e_{n} \text{e}^{(i\mathbf{k}\cdot\mathbf{r})}\frac{d^{3}\mathbf{k}}{(2\pi)^{3}} +\sigma_{ij}^{\mathrm{appl}}M_{ij}(\alpha),
\end{eqnarray}
where the Fourier transform of the stress $ \hat{\sigma}_{mn}^{0}(\mathbf{k})$ is defined by (\ref{F_stress}).

The set of equations (\ref{FDislocation2}) - (\ref{SlipEvolution}),  (\ref{SlipEvolution2}), (\ref{TauExternal}) are calculated in each step of the simulation for all slip systems, then the plastic shear functions $\gamma_{\alpha}$ and dislocation densities are inserted in the kinetic equation (\ref{KineticEq}).

\section{Numerical simulation}
\label{Section_Numerical_simulation}

\subsection{Choice of parameters}

For the numerical investigation we chose a Fe-30$\%$Ni alloy, where the occurring lenticular-type martensite is characterized by the mixture of  twinning and dislocation structure.

The material parameters and phase-field model parameters used in the simulations
are listed in Tables \ref{Table1}, \ref{Table2}.

The reason of the choice of the discretization size is that we first chose the relation between the interface width and the maximal capillary 
length $d_0=\gamma /(a_1 E_0)=1.3\times10^{-11}$ m. That implies that $\xi/d_0<200$, which is a requirement of the asymptotic limit in the phase-field model \cite{Karma98}.
In the simulations we used  the phase-field equations in dimensionless form by measuring length in units $l_0$, time in units of $\tau_0$ ( it can be estimated from the experimental transformation rate for the MT of order $100$ ms$^{-1}$) and the energy in units of $E_0$, which is the typical strain energy of MT. 
Then the mobility of the interface is calculated as $\mathcal{M}=1/(\tau E_0)$.

\begin{table}
\caption{\label{Table1} Material and processing parameters used in the simulation}
\begin{center}
\begin{tabular}{|l|l|}
 \hline
Parameter & Value \\ \hline
Temperature of MT & $ T_M=400$ K\\
Interface energy  & $\sigma=1.9\times10^{-2}$ J m$^{-2}$ \cite{Artemev2002}\\
Latent heat       & $Q_M=3.5\times10^8 $ J m$^{-3}$  \cite{Artemev2001}  \\
 Initial disl. density  & $\rho_{I,0}=10^{10}$ m$^{-2}$ \\
External applied stress & $\sigma_{11}^\mathrm{appl}=\sigma_{33}^\mathrm{appl}=-10$ GPa\\
Shear modulus    & $G=28 $ GPa\\
Poisson coefficient & $\nu=0.374$\\
Material constants  & $c_1=0.18$\\
                  & $c_2=5.0$\\
                  & $c_3=5.0$\\
                  & $c_4=8.0\cdot10^6$ m$^{-1}$\\
                  & $c_5=10$\\
                  & $c_7=1\times10^{7}$ m$^5$s$^{c_8}$\\
                  & $c_8=0.3$\\
                  & $c_9=5\times10^{-10}$ m$^2$ \\
                  & $c_{10}=0.1$\\
Attack frequency  & $\nu_{0}=10^{10}$ s$^{-1}$\\
Activation energy & $Q_{\mathrm{slip}}=2.3\times10^{-19} $ J\\
Activation energy & $Q_{\mathrm{bulk}}=2.4\times10^{-19} $ J\\
Lattice constant & $a_0= 3.59\times10^{-10}$ m \\
\hline
\end{tabular}
\end{center}
\end{table}

\begin{table}
\caption{\label{Table2} Phase-field model parameters used in the simulation}
\begin{center}
\begin{tabular}{|l|l|l|}
 \hline
Parameter & Value & Units \\ \hline
Length scale, $l_0$	    & $1.3\times10^{-9} $ & m\\
Time scale, $\tau_0$  	    & $3.3\times10^{-10}$ &s\\
Energy scale, $E_0$ & $3.07\times10^9$ & J m$^{-3}$\\
Discretization size, $\Delta x$  & $1$ &$l_0$\\
Time step, $\Delta t$ & $0.125$ & $\tau_0$\\
Interface width, $\xi$  &$1.5$ & $ l_0$\\
Gradient coefficient, $K$ & $0.0152$ & $E_0l_0^2$\\
Double-well potential coefficient, $H$ & $0.0067 $ & $E_0$\\
Landau energy parameter, $a_2$ & $0.0134 $ & $E_0$\\
Interface mobility, $\mathcal{M}$ & $1 $ & $(\tau_0 E_0)^{-1}$ \\
Undercooling, $\Delta f$ & 0.06 & $E_0$\\
\hline
\end{tabular}
\end{center}
\end{table}

 For these parameters the Landau energy parameter $a_2=2\sigma/(a_1\xi)$ in the chemical energy function (see Appendix A) equals to  $a_2=0.0134\,E_0$.  The estimation of the gradient coefficient in eq.~(\ref{KineticEq}) yields $K=\sigma \xi/(a_1)=0.0152\,E_0l_0^2$. For the comparison with work \cite{Zhang2007} the gradient coefficient is in the same order ($K=0.01624\,E_0l_0^2$) but the Landau energy parameter is ten times larger ($a_2=0.312\,E_0$). The interface width used in the simulation was very much smaller  than the chosen discretization size $\xi=\sqrt{K/H}=0.32\, l_0$. This means that the increasing parameter $a_2$ results in a decreasing interface width. In the simulation we used the parameters listed in Table \ref{Table1}. For comparison we also carried out the simulations with $a_2=0.312\,E_0$ ($\xi=0.32\, l_0$) to show the influence of this parameter on the annihilation term.

For three variants of the MT we used the Bain transformation matrices: 
\begin{eqnarray}
\varepsilon_{11}^0=0.1322, \varepsilon_{22}^0=0.1322, \varepsilon_{33}^0=-0.1994, \text{variant 1};\nonumber\\
\varepsilon_{11}^0=-0.1994, \varepsilon_{22}^0=0.1322, \varepsilon_{33}^0=0.1322, \text{variant 2};\nonumber\\
\varepsilon_{11}^0=0.1322, \varepsilon_{22}^0=-0.1994, \varepsilon_{33}^0=0.1322, \text{variant 3}.\nonumber
\end{eqnarray}

In the fcc lattice of the austenitic matrix 24 variants of the dislocation slip
were chosen with slip planes of type $\lbrace111\rbrace_{A}$ and  slip directions of type $\langle101\rangle_{A}$. For the calculation of the Burgers vector we used $b=a_0/\sqrt{2}$, where  $a_0$ is the lattice constant.

\subsection{Numerical investigation of the dislocation inheritance} \label{Section_Prob1}

 In this section we present the simulation results for the case of the pre-deformed austenite where the initial dislocation density  for all slip systems ($\rho_{I,0}$ in Table \ref{Table1}). A nucleus containing 3 martensitic variants are generated in a system of size $64\times 64\times 64\, \Delta x$.  Periodic boundary conditions are applied to the systems for all fields.

An external force of 1 GPa is imposed in direction $[\bar{1}0\bar{1}]$. This force suppresses one martensitic variant and promotes two other variants (variants 1 and 2), which produce a twinning structure.  The dislocation density evolves during the simulation due to the applied stress and the stress from the MT.

 The probability of dislocation inheritance from the austenitic matrix to the martensitic phase is calculated in the simulation as a ratio between the averaged dislocation density with and without the annihilation term in eq.~(\ref{DensEvolution}) 
\begin{equation}
\mathcal{P}_\alpha(t)=\frac{\bar{\dot{\rho}}_{I,\alpha}}{\bar{\dot{\rho}}_{I,\alpha} +\overline{c_9\rho_{I,\alpha}\rho^f_M\dot{\eta}}},\label{Prob1}
\end{equation}
where $\dot{\eta}=\sum_{p=1}^\mu\dot{\eta}_{p}$ and $\bar{x}$ identify a space averaged variable over the interface between martensitic and austenitic phases
\begin{equation}
\bar{x}=\frac{\int_0^{V_{box}}x\eta\phi_AdV}{\int_0^{V_{box}} \eta\phi_A dV}.
\end{equation}
Thus, we consider only the dislocations which can take part in the annihilation process at the MT front.
 Note that from the probability $\mathcal{P}_\alpha(t)$ the contribution of the annihilation term to the total dislocation density is derived as $1-\mathcal{P}_\alpha(t)$.

During the simulation we monitor the ratio between the mean dislocation densities in the martensite (all variants) and austenite

\begin{equation}\label{Prob}
\mathcal{R}_\alpha(t)=\frac{\bar{\rho}_{I,\alpha}\vert_M}{\bar{\rho}_{I,\alpha}\vert_A}.
\end{equation}

\begin{figure}
\begin{centering}
\includegraphics[scale=0.45]{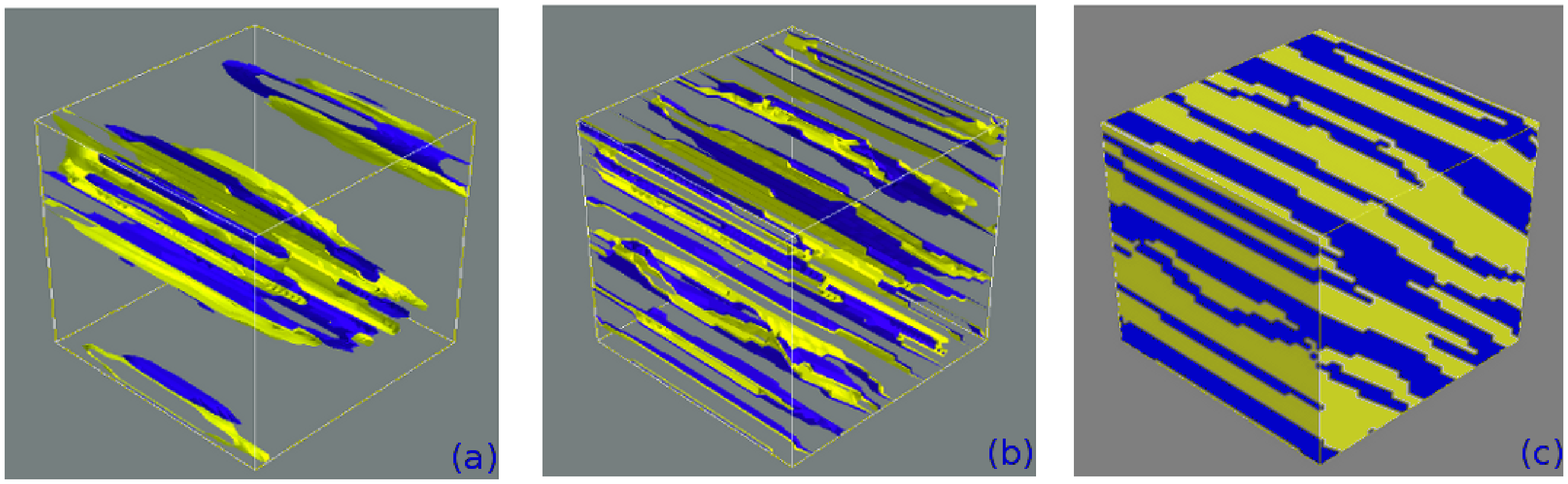}  
\caption{\label{Fig2}
3D images of simulated microstructures during MT (test 1) for the  box of size $38 \, \Delta x$ with $\xi=1.5\, l_0$ and 1 nucleus at 2500 (a) and 5000 (b) time steps; the surface view at 5000 time steps (c).  The yellow and blue colors define two martensitic variants.}
\end{centering}
\end{figure}

\begin{figure}
\begin{centering}
\includegraphics[scale=0.45]{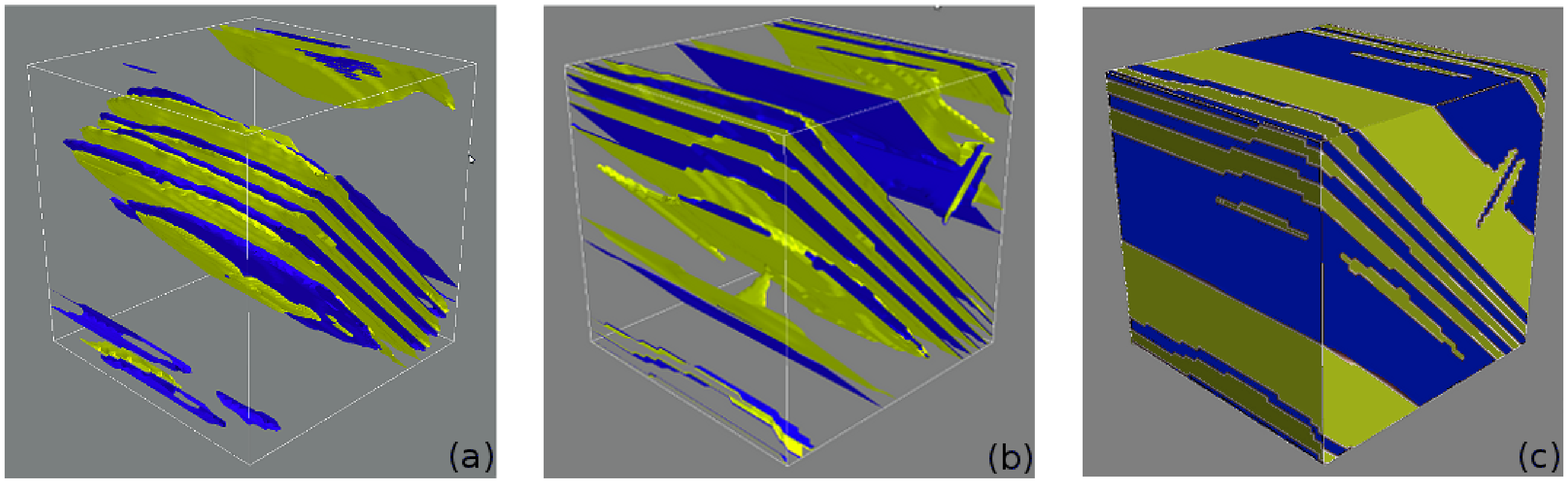}  
\caption{\label{Fig3}
3D images of simulated microstructures during MT (test 2) for the  box of size $64 \, \Delta x$ with $\xi=1.5\, l_0$ and 1 nucleus at 2500 (a) and 5000 (b) time steps; the surface view at 5000 time steps (c).  The yellow and blue colors define two martensitic variants.}
\end{centering}
\end{figure}

\begin{figure}
\begin{centering}
\includegraphics[scale=0.45]{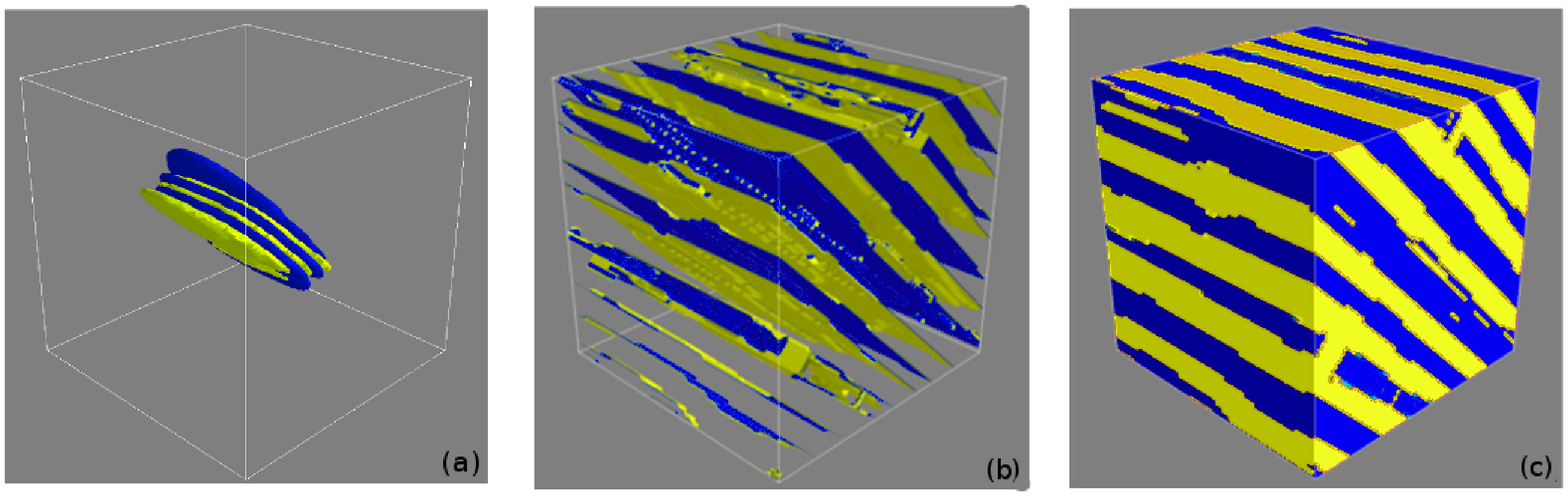}  
\caption{\label{Fig3a}
3D images of simulated microstructures during MT (test 3) for the  box of size $64 \, \Delta x$ with $\xi=0.32\, l_0$ and 1 nucleus at 2500 (a) and 5000 (b) time steps; the surface view at 5000 time steps (c).  The yellow and blue colors define two martensitic variants, white color defines the austenite.}
\end{centering}
\end{figure}

\begin{figure}
\begin{centering}
\includegraphics[scale=0.45]{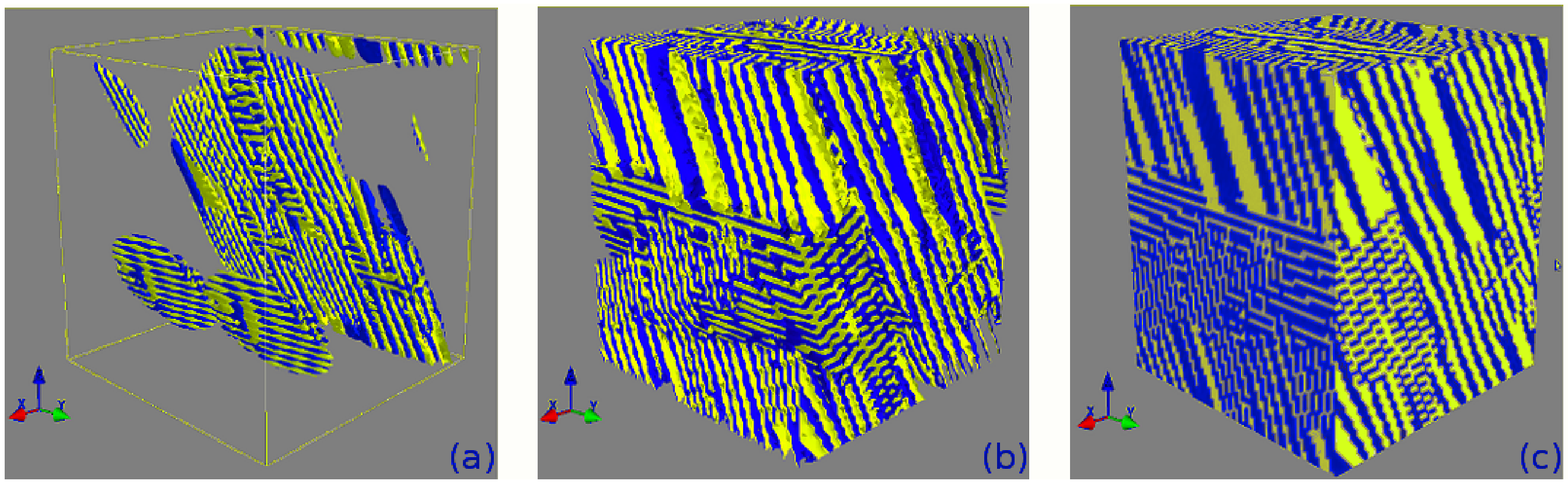}  
\caption{\label{Fig4}
3D images of simulated microstructures during MT (test 4) for the box of size $64 \,\Delta x$  with $\xi=0.32\, l_0$ and 6 nucleus at 2500 (a) and 5000 (b) time steps; the surface view at 5000 time steps (c). The yellow and blue colors define two martensitic variants. }
\end{centering}
\end{figure}

 The simulated microstructure during the MT at 1500, 2500 and 5000 time steps is shown in Figs. ~\ref{Fig2}--\ref{Fig4} for four tests. In test 1 and test 2 (Figs.~\ref{Fig2} and ~\ref{Fig3}) we simulated the microstructure with the interface width $\xi=1.5\, l_0$ in boxes of size $38\,\Delta x$ and $64\,\Delta x$ with one initial nucleus. The microstructure consists of twinned martensitic plates, which belong to one martensitic lath. The size of the system influences the number and the thickness of the growing plates. It can be seen that for the smaller system size the number of the martensitic plates is larger and their thickness is evenly distributed in the volume. For the smaller system size the thickness of martensitic plates becomes 2-3 times larger at the end of the growth. 

 In test 3 and 4 (Fig.~\ref{Fig3a} and \ref{Fig4})  the interface width was chosen $\xi=0.32\, l_0$ in a simulation box of size $64\,\Delta x$. In test 3 (Fig.~\ref{Fig3a}) one initial nucleus is used similar to tests 1 and 2.  In test 4 (Fig.~\ref{Fig4})  6 nuclei of the martensitic phase were inserted randomly with the mean time  interval equal to 100 time steps and the coordinate system was rotated to [$\bar{1}\bar{1}2$],[111],[$\bar110$] for the comparison to Ref.~\cite{Zhang2007}.  Simulations show that for the small interface width $\xi=0.32\, l_0$  the MT can not be completed at the chosen undercooling. To reach the full martensitic transformation the simulation in test 4 was carried out at the increasing undercooling from $\Delta f$ to $3 \Delta f$ for 2500 simulation time steps that corresponds to the cooling rate $2\cdot 10^8$ K/s. In comparison to Fig.~\ref{Fig3} the smaller interface width $\xi=0.32\, l_0$  gives more branched microstructure that is similar to the microstructure in Ref.~\cite{
Zhang2007}. But in this case the 
phase-field model has maximal 2 discretization points on the interface, while for $\xi=1.5\, l_0$ the phase-field model gives minimal 4-5 discretization points. We suggest to use the interface width $\xi=1.5 l_0$, which is a relevant value for the diffuse interface in the phase-field model.

\begin{figure}
\begin{centering}
\begin{tabular}{cc}
\includegraphics[scale=0.4]{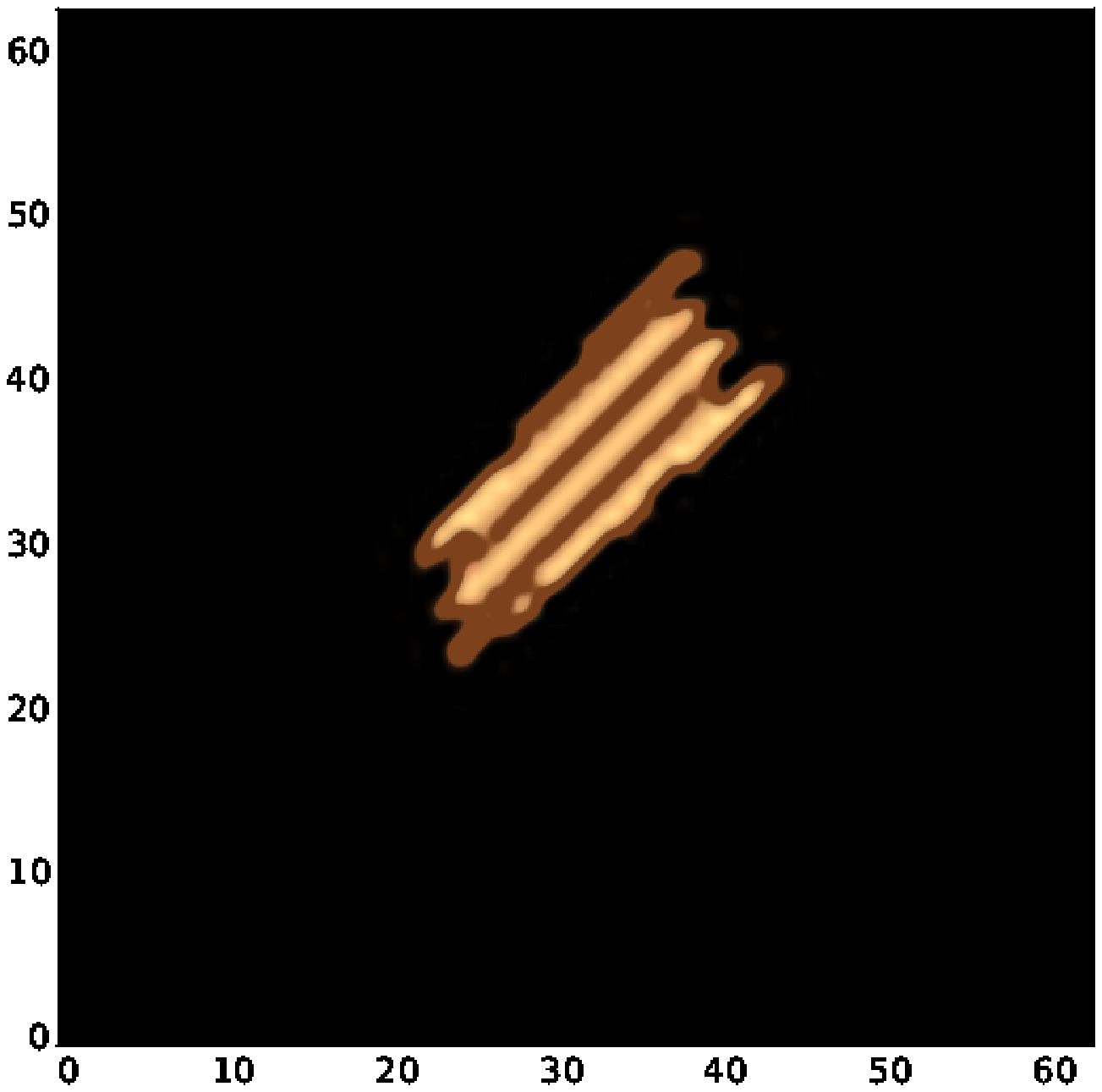} & \includegraphics[scale=0.4]{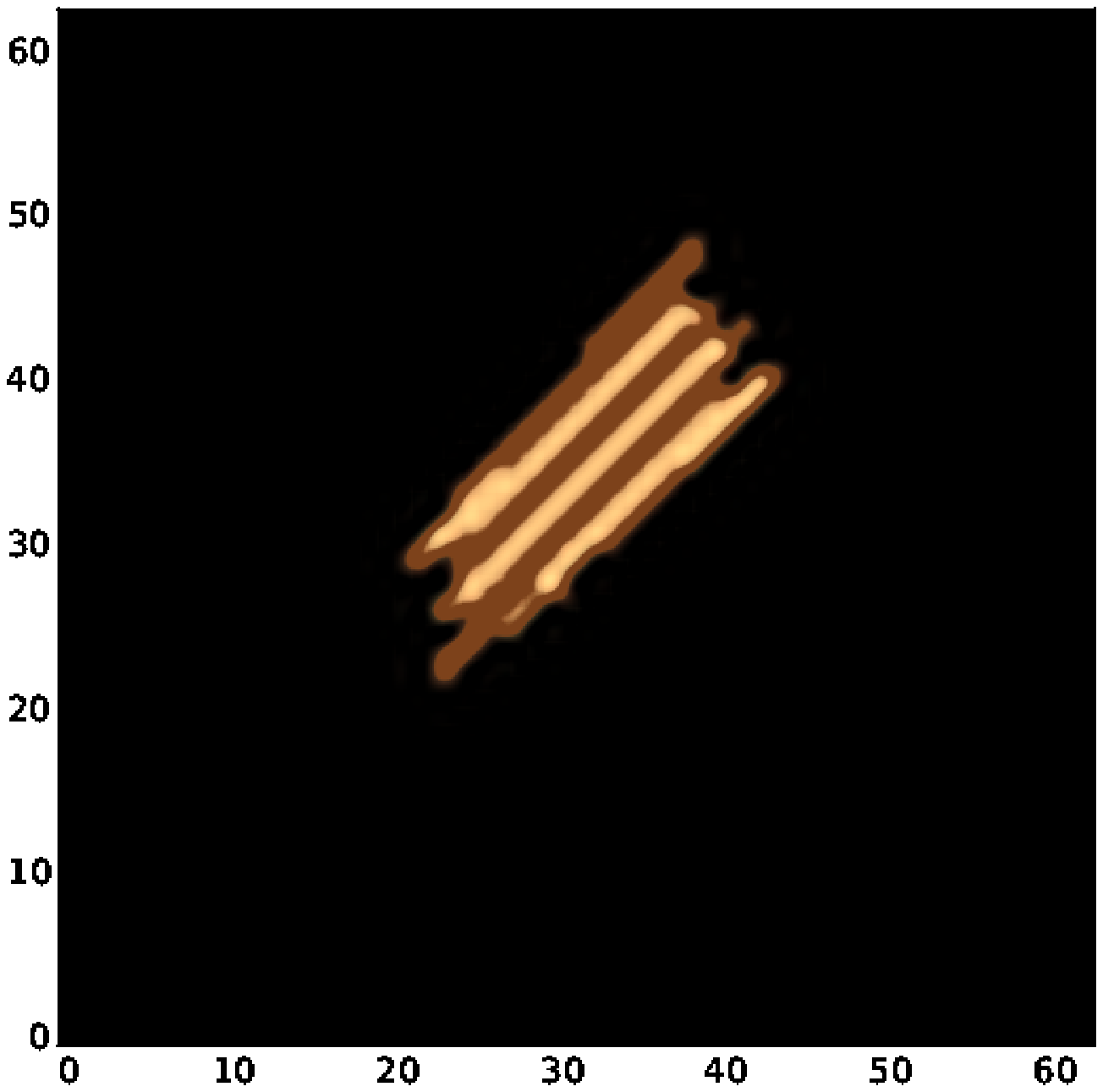} \\
\includegraphics[scale=0.4]{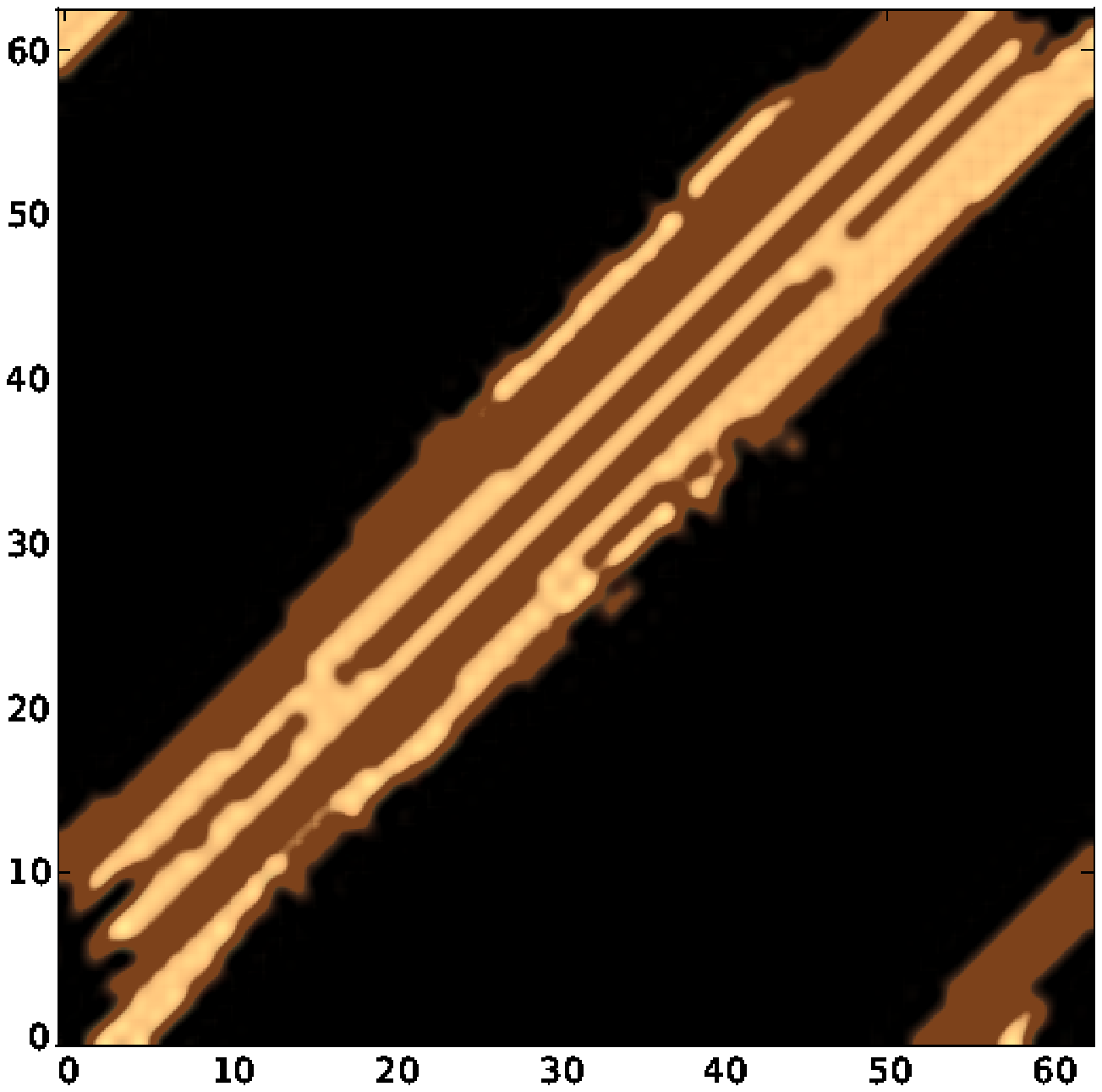} & \includegraphics[scale=0.4]{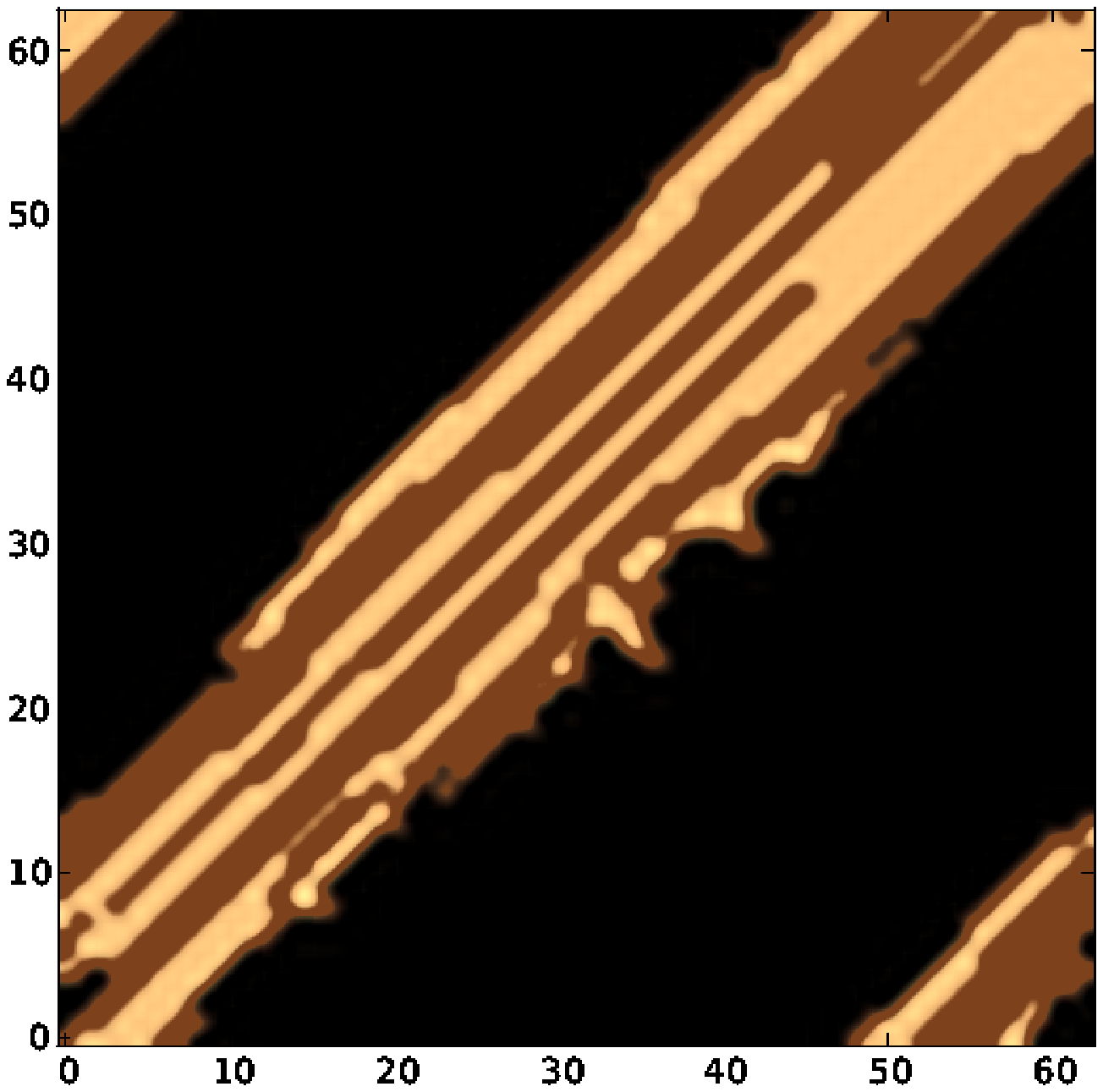} \\
\includegraphics[scale=0.4]{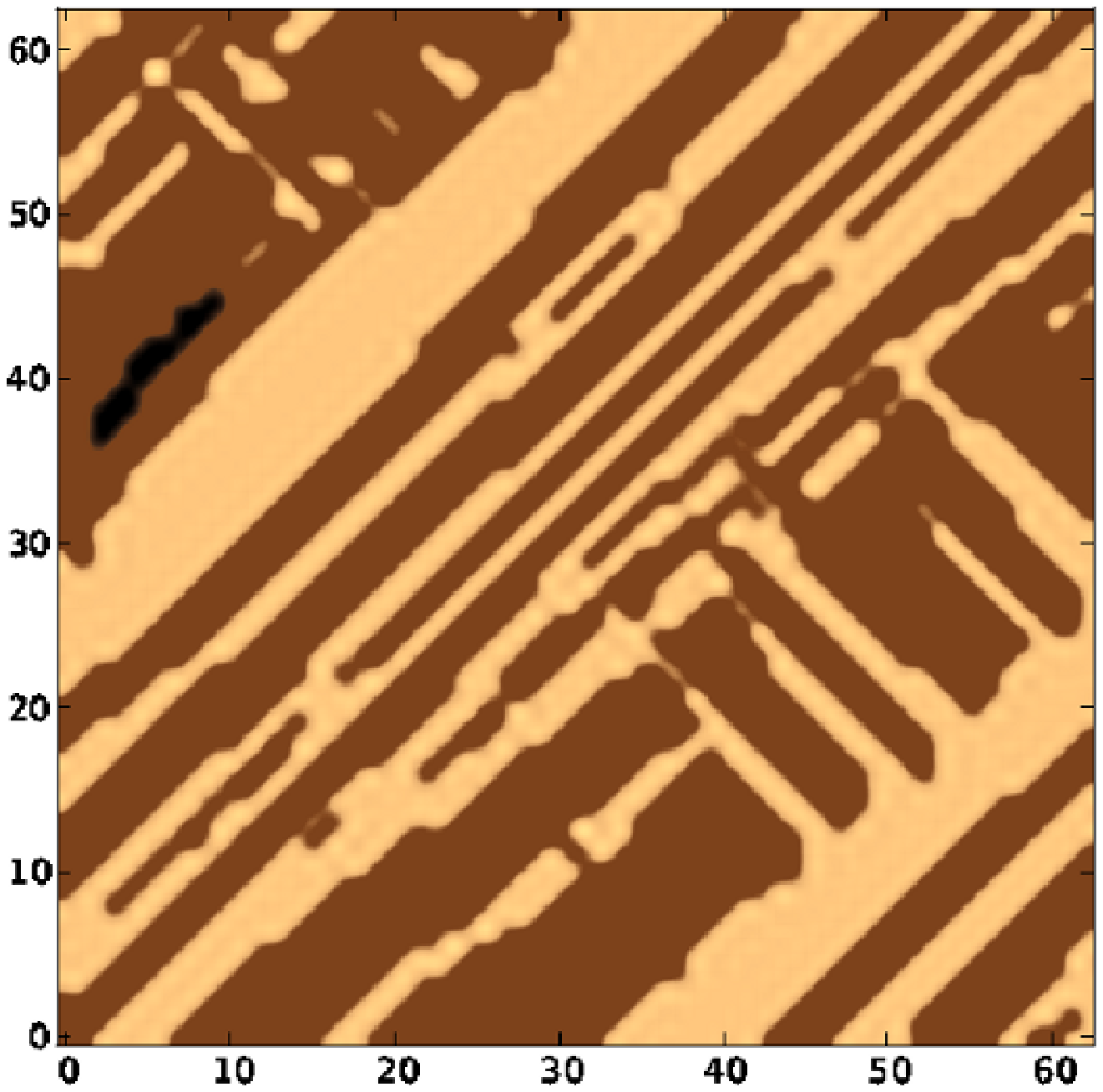} & \includegraphics[scale=0.4]{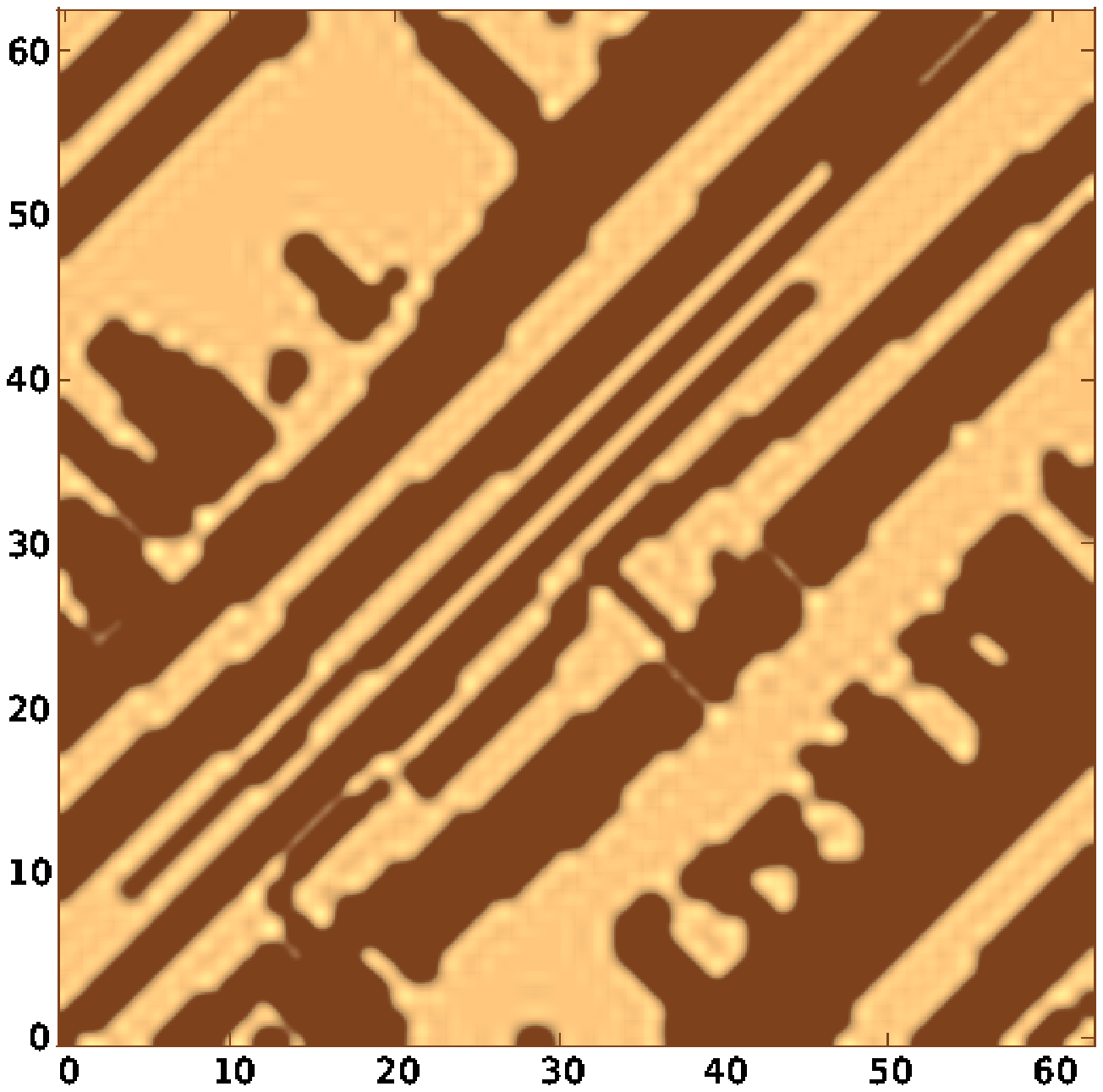}\\
\end{tabular}
\caption{\label{Fig5}
Simulated microstructure evolution during MT in 2D cross-sections of a 3D box of size $64 \,\Delta x$ along [001] axis. First column contains the microstructures  without plastic effects at the time 100$\tau_0$,150$\tau_0$ and  350$\tau_0$, second column contains the microstructures without plastic effects at the time 100$\tau_0$,150$\tau_0$ and  250$\tau_0$.   The yellow and brown colors define two martensitic variants 1, 2 and black color defines the austenite.}
\end{centering}
\end{figure}

 The 2D sections ($x-y$-directions) of the microstructure are presented in Fig.~\ref{Fig5} for the system size $64\,\Delta x$. The first column contains the time evolution of the microstructure without plastic effects at the time 100$\tau_0$,150$\tau_0$ and 350$\tau_0$, the second volume contains the time evolution of the microstructure without plastic effects at the time 100$\tau_0$,150$\tau_0$ and  250$\tau_0$. The microstructure does not distinguish between the cases with and without annihilation term. 

\begin{figure}
\begin{centering}
\begin{tabular}{cc}
\includegraphics[scale=0.25]{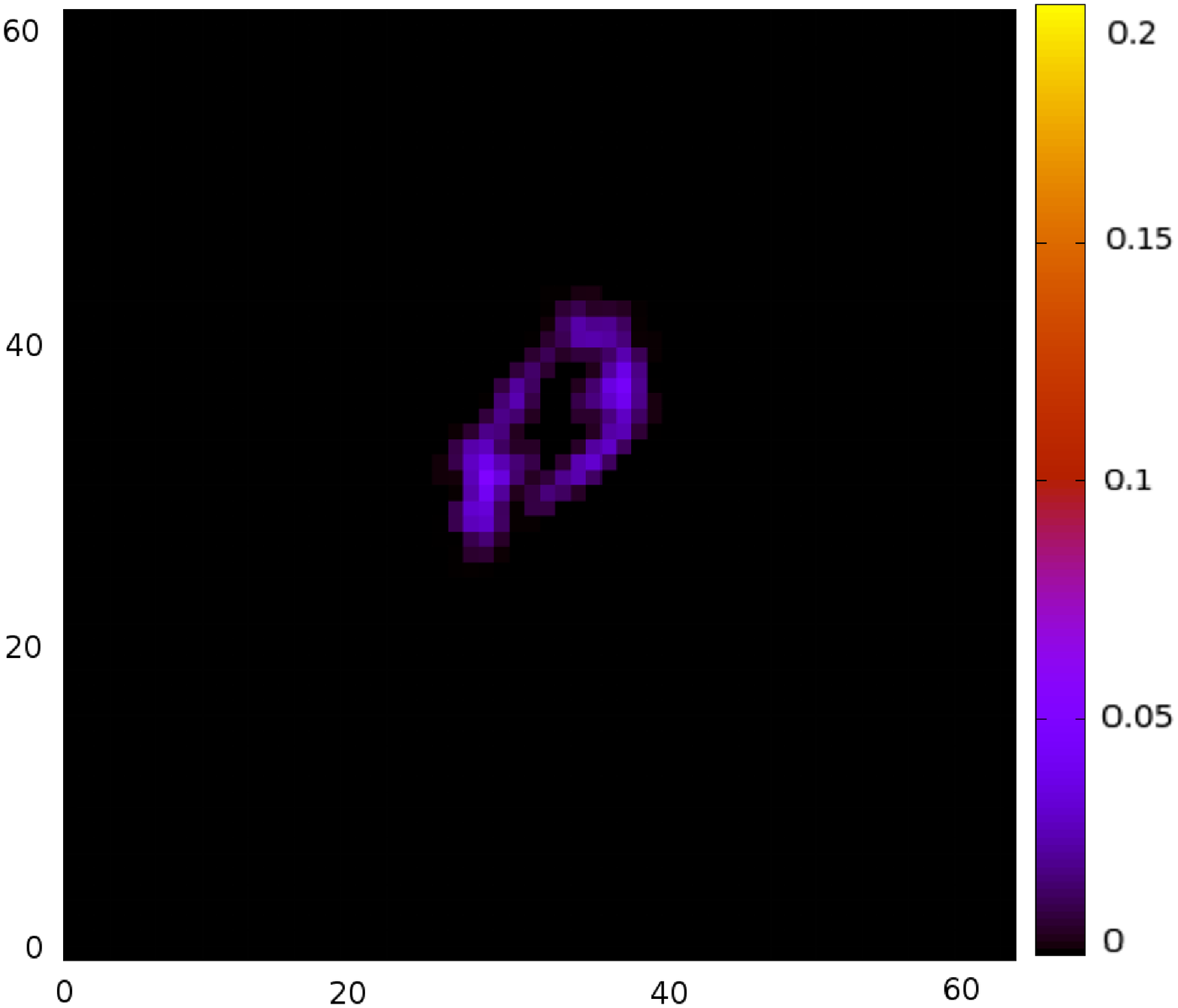} & \includegraphics[scale=0.25]{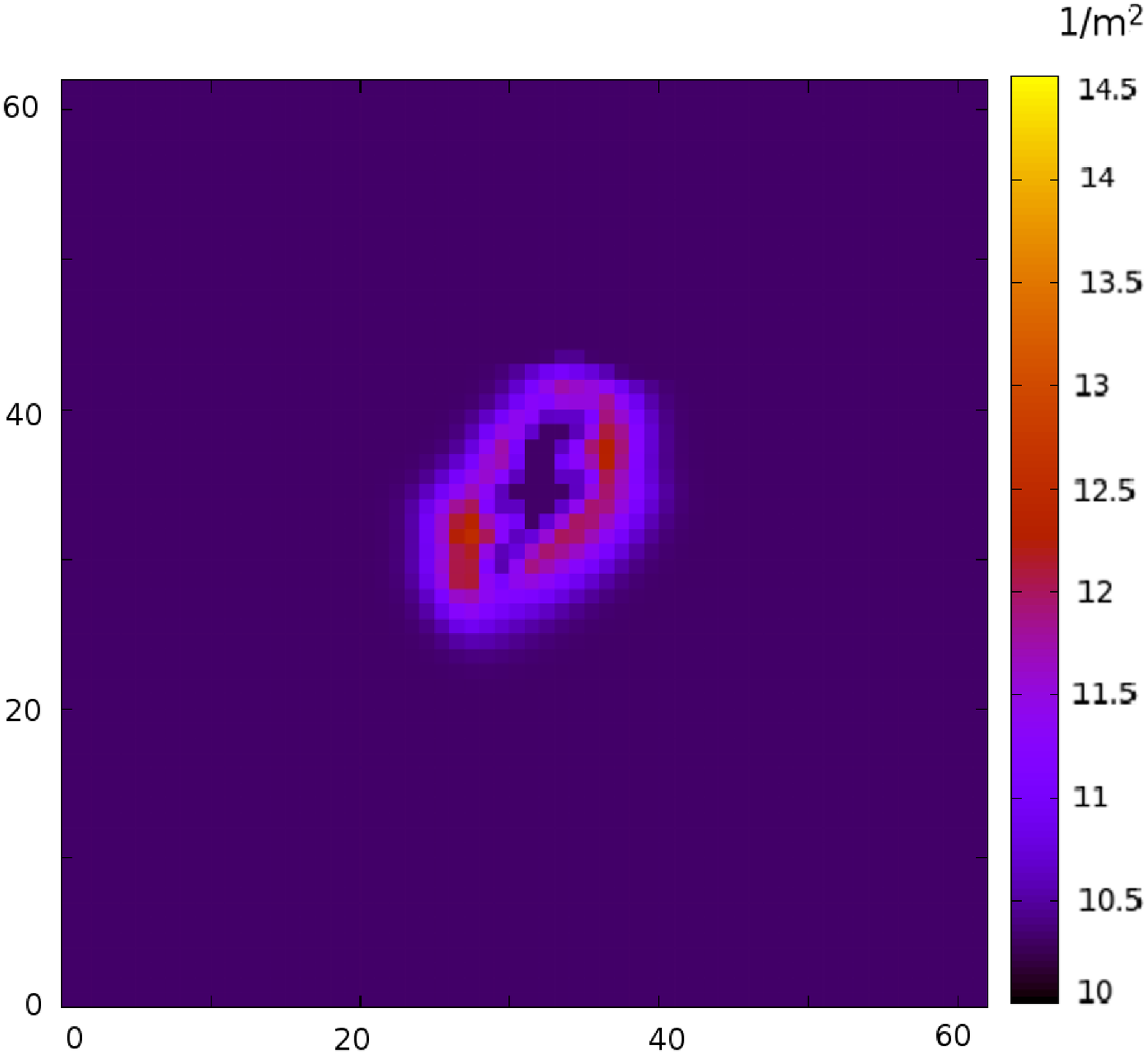} \\
\includegraphics[scale=0.25]{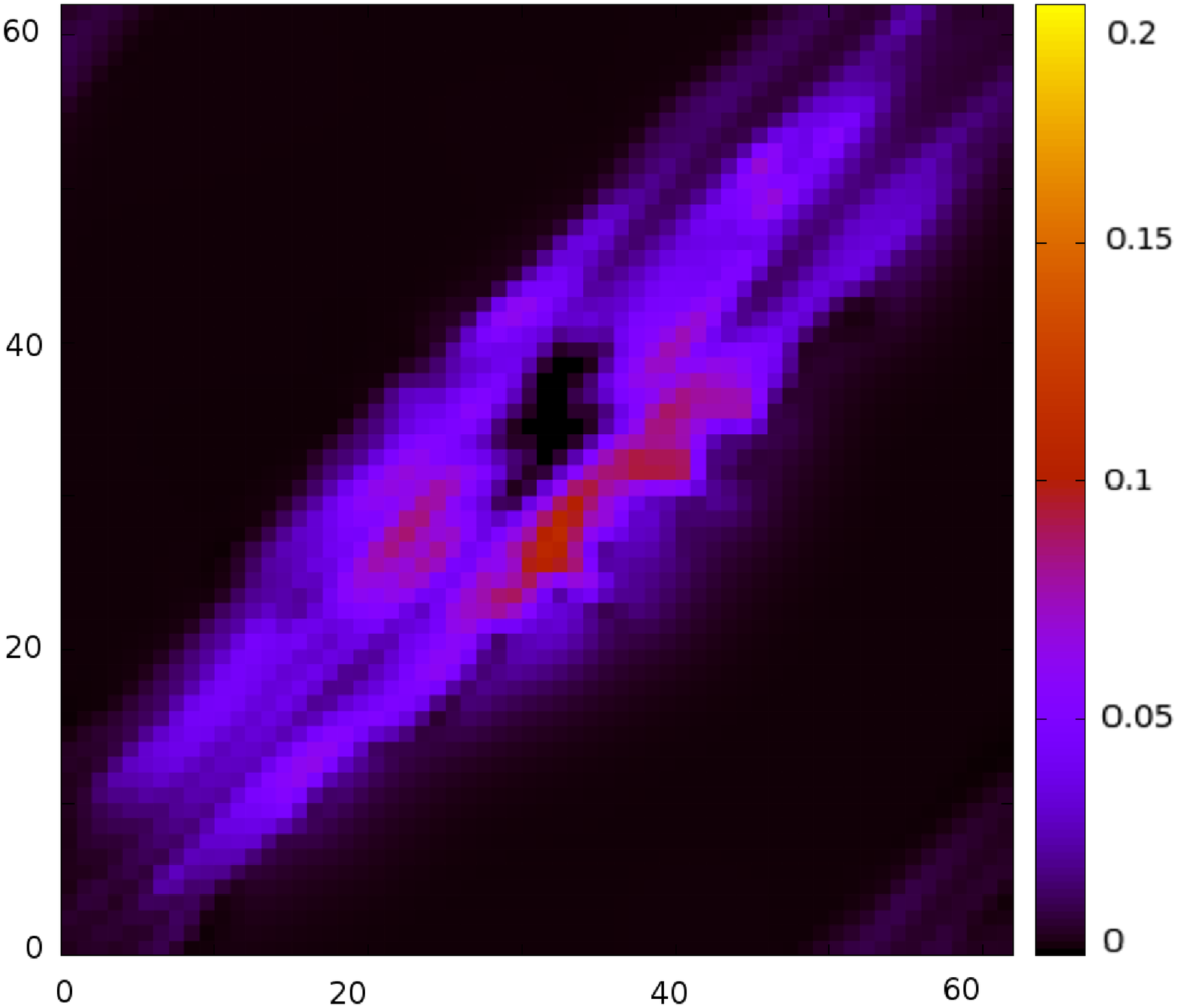}& \includegraphics[scale=0.25]{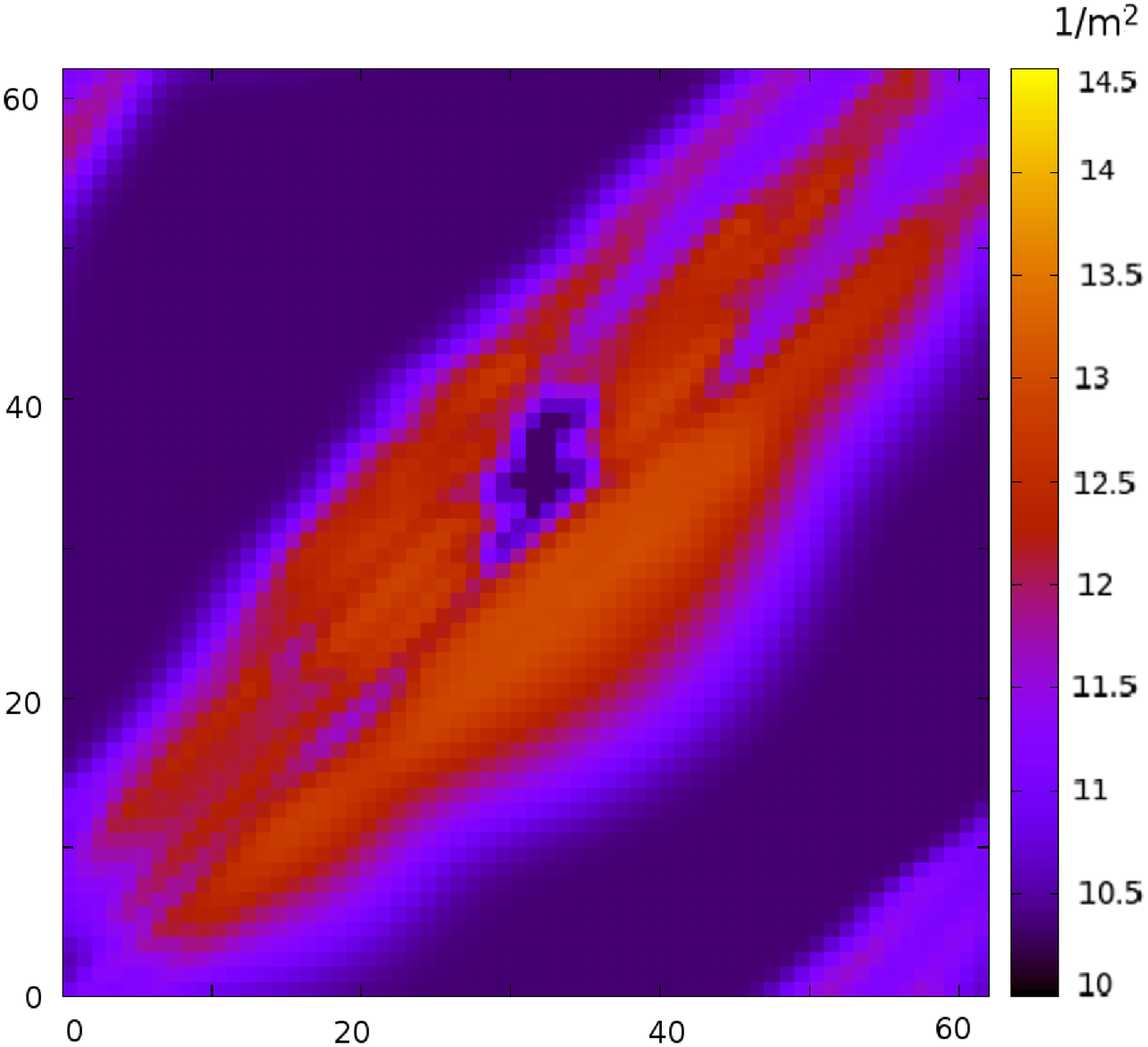} \\
\includegraphics[scale=0.25]{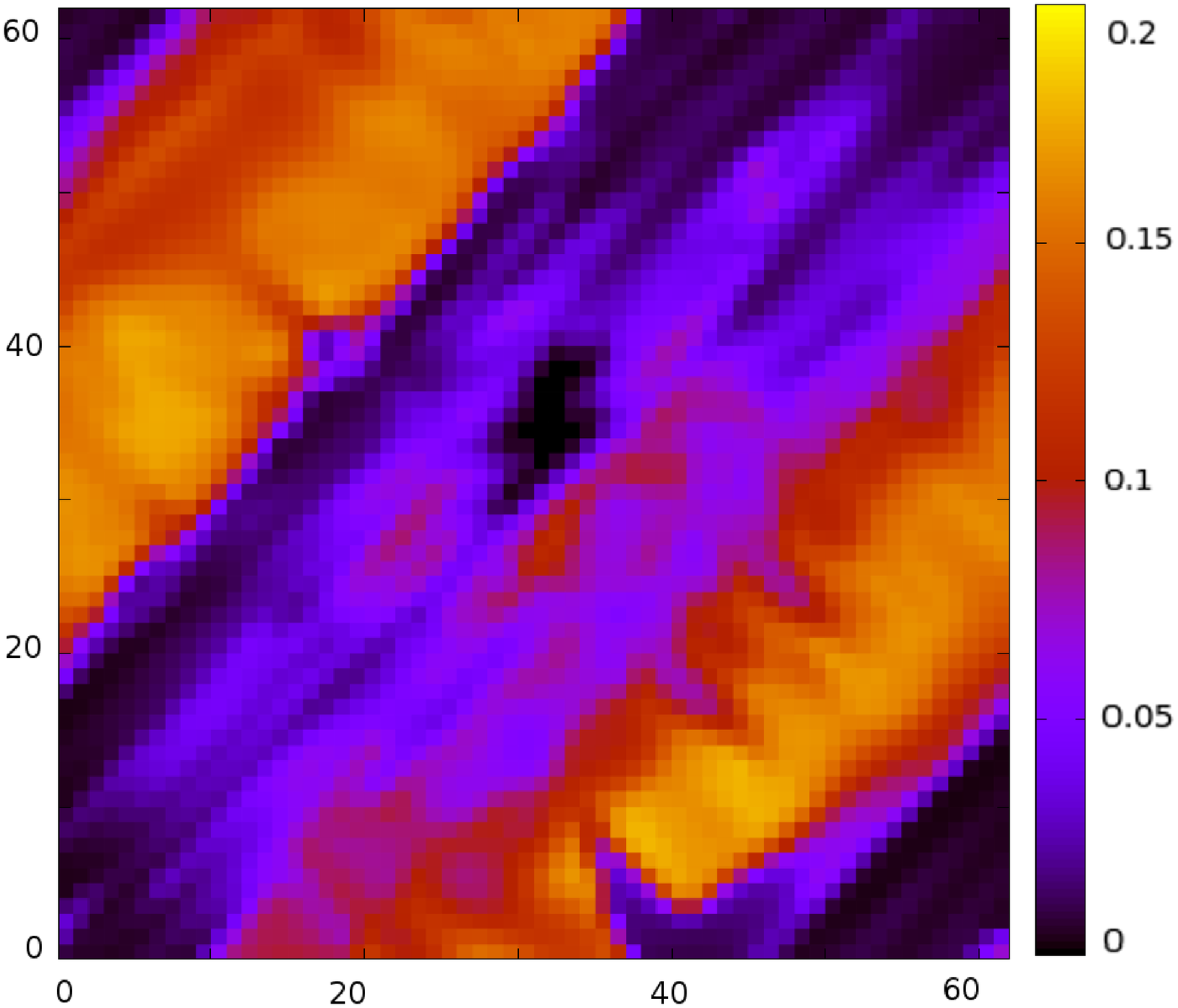} & \includegraphics[scale=0.25]{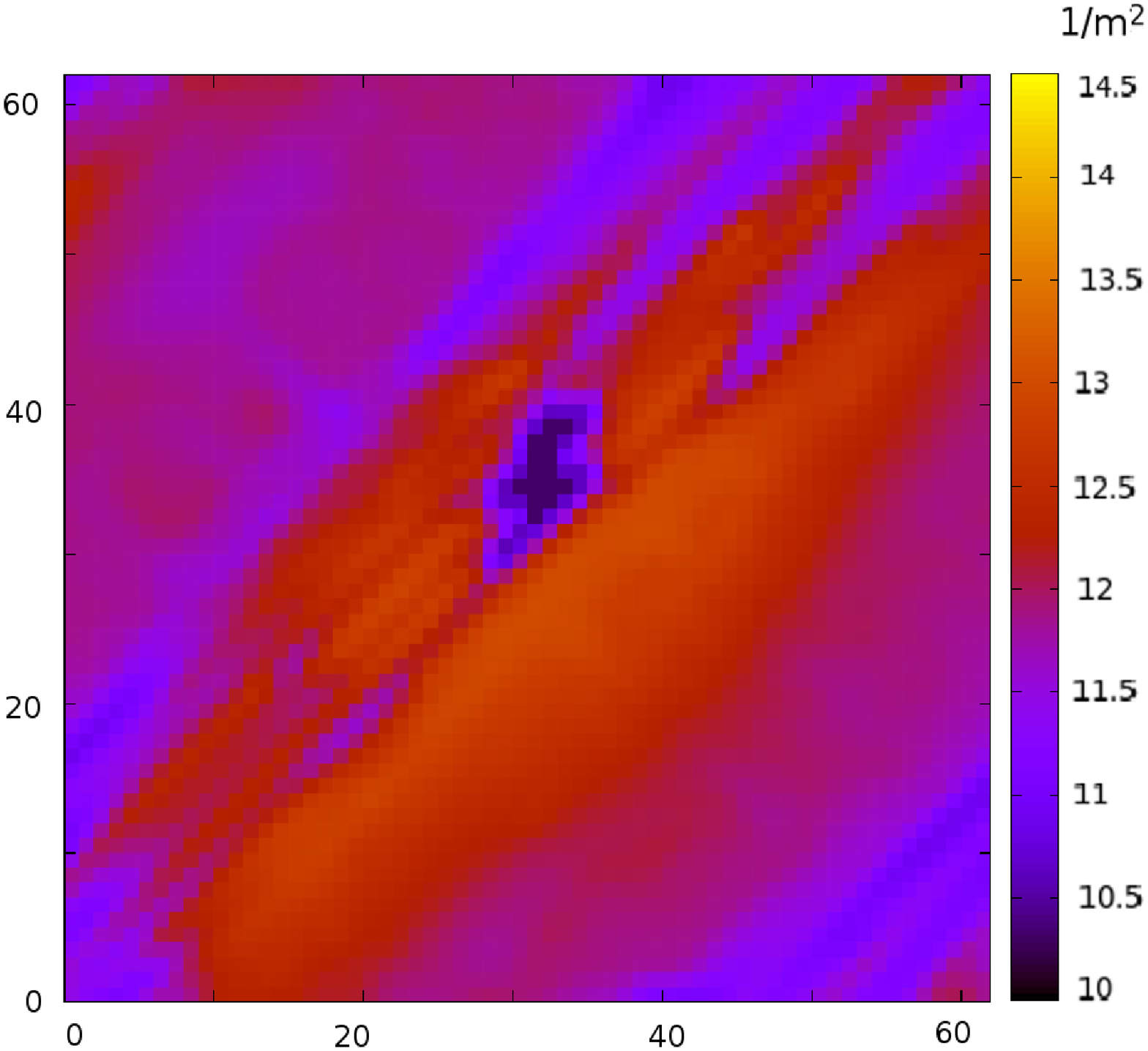} \\
\end{tabular}
\caption{\label{Fig6}
Simulated plastic strain (left)  and the dislocation field (right) during MT in 2D cross-sections of a 3D box of size $64 \,\Delta x$ along [001] axis with plastic effects at the time 75$\tau_0$,150$\tau_0$ and  250$\tau_0$. The summary dislocation field is given in logarithmic scale. }
\end{centering}
\end{figure}

 The 2D sections ($x-y$-directions) of the summary plastic strain, $\gamma_\alpha$, and the summary dislocation density field, $\phi^d$, (in logarithmic coordinates) are presented in Fig.~\ref{Fig6}. They corresponds to the microstructure in Fig.~\ref{Fig5}. It can be seen that the dislocation density increases first near the transformation front in the earlier stages of MT and strongly increases in the full austenitic volume at the end of the MT. 

\begin{figure}
\begin{centering}
\begin{tabular}{cc}
\includegraphics[scale=0.4]{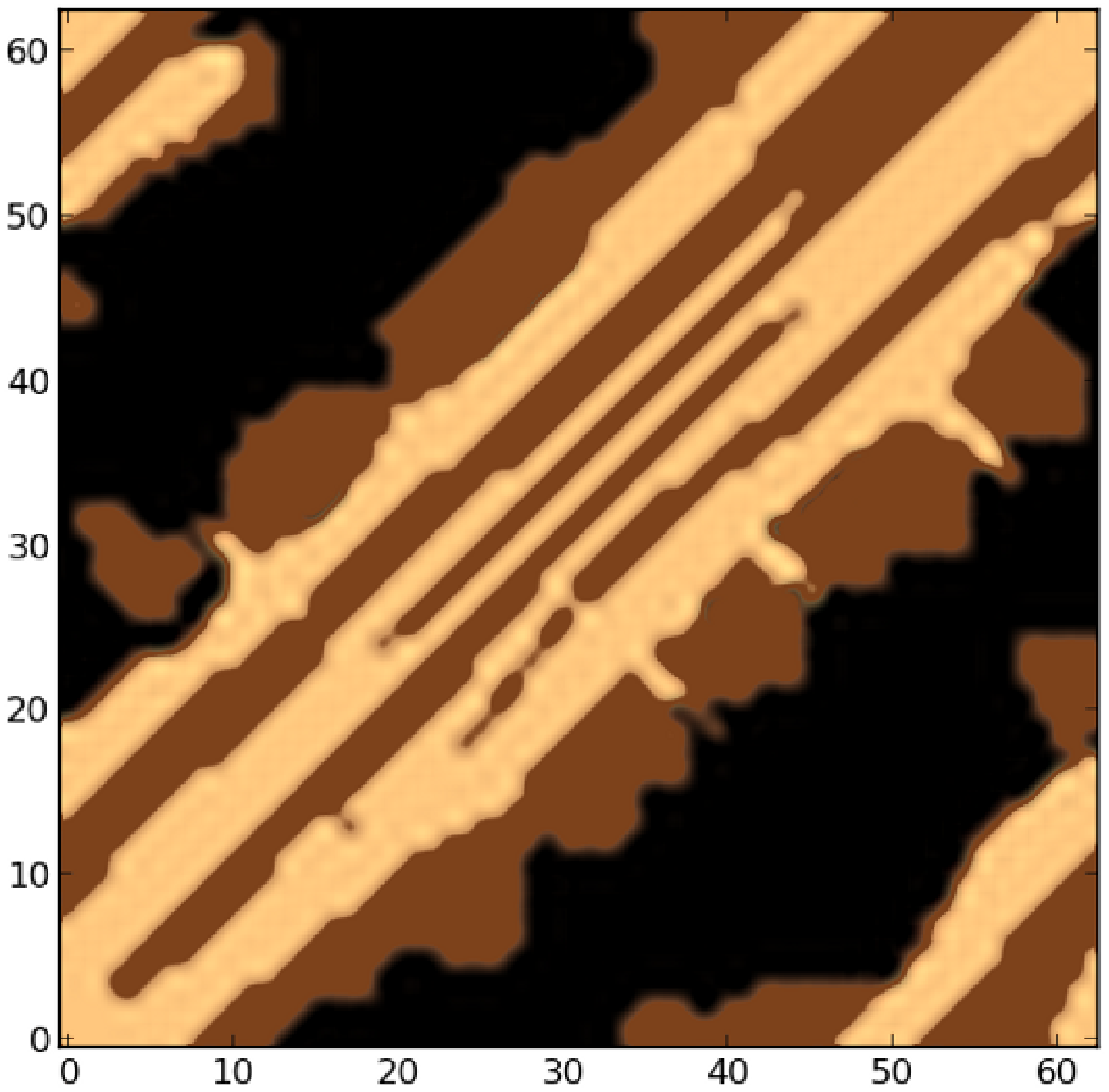} & \includegraphics[scale=0.25]{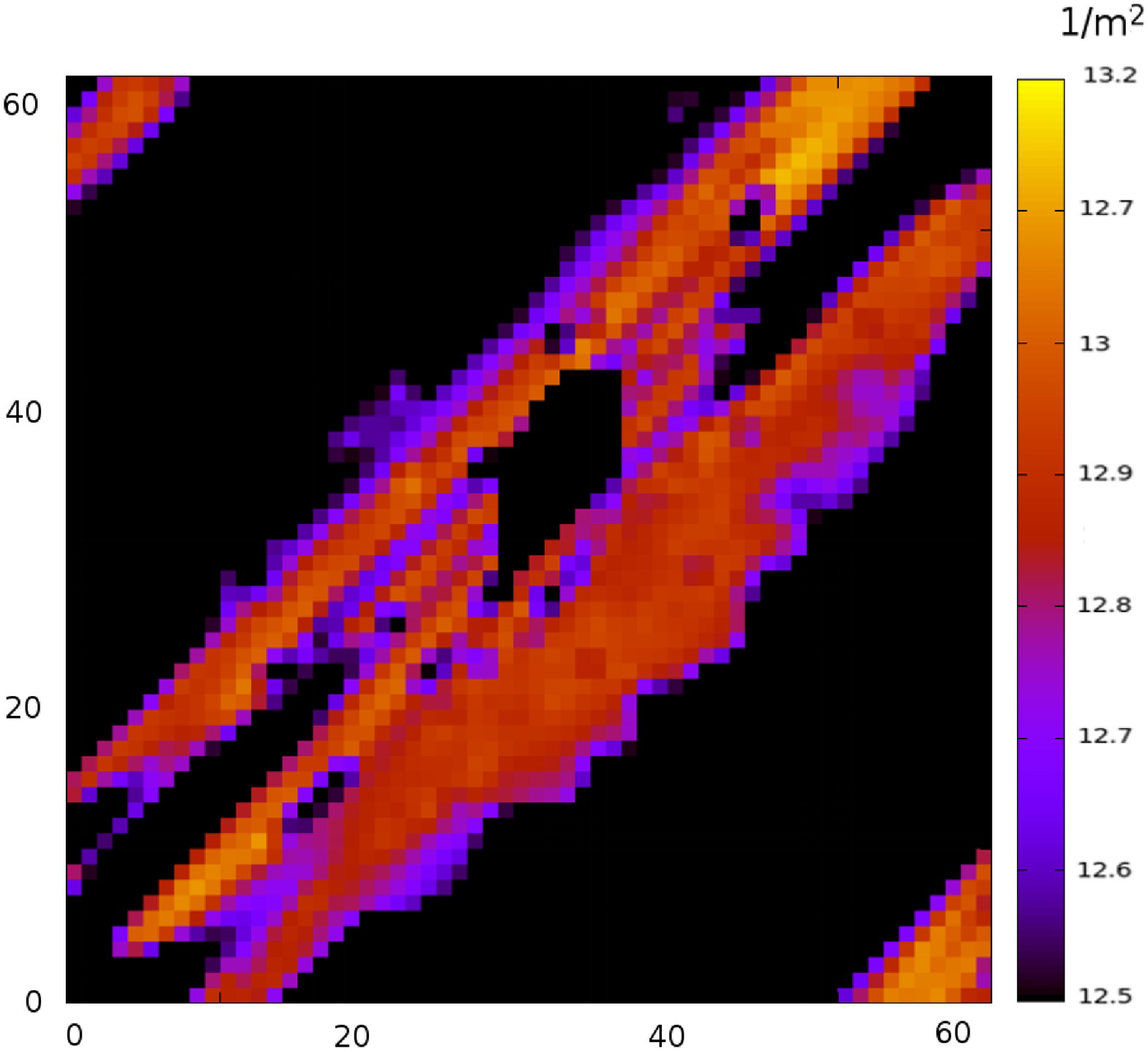} 
\end{tabular}
\caption{\label{Fig6ADD} 
The comparison of the microstructure (a) and the summary dislocation density field (b) for two slip systems (1) and (2) (Table \ref{Table3}) at the time 200$\tau_0$.}
\end{centering}
\end{figure}

 For the analysis we chose two slip systems as examples 
of low and high interaction stresses between the martensite and the dislocation field. As a characteristic parameter of this interaction a maximum interaction stress can be proposed  
\begin{equation}\label{Stress_free}
\tau_{\mathrm{inter}}(\alpha,p)=c_{ijkl} M_{ij}(\alpha) \varepsilon_{kl}^0(p), 
\end{equation}
which influences the kinetics of the MT and the evolution of the plastic strain and the dislocation density due to the phase transformation (see eq.~\ref{TauExternal}). It is obviously that the interaction stress between the martensitic plates and the plastic strain depends on the orientation relationship between the corresponding martensitic variants and the slip systems.
 The full list of the interaction stresses between 12 slip systems  and two martensitic variants is presented in Table \ref{Table3}.  As an example for the visualization and the qualitative analysis we chose two typical slip systems (1) and (2) being indicated in Table \ref{Table3} and having different summary interaction stresses (see last column in Table \ref{Table3}).  The summary dislocation density field  of two slip systems (1) and (2) is shown in   Fig.~\ref{Fig6ADD} in comparison to the microstructure. It can be observed that the dislocation density  is larger in the martensitic variant 2 (brown color) than in variant 1. It is because in the variant 1 the interaction stress is zero for the slip system (1) and in the variant 2 the interaction stress is not zero for both chosen slip systems.

\begin{table}
\caption{\label{Table3} Interaction stress between the dislocation slip systems and martensitic variants. The slip systems used in the simulation are underlined.}
\begin{center}
\begin{tabular}{|l|l|l|l|}
 \hline
Slip system & Mart. variant 1& Mart. variant 2 & sum\\ \hline
$\underline{(111)[\bar{1}10]}$ (1) & 0 &0.0736807 & 0.0736807\\
$(11\bar{1})[\bar{1}10]$ & 0 &0.0736807& 0.0736807\\

$(111)[0\bar{1}1]$ & -0.0736807 &0 &-0.0736807\\
$(\bar{1}11)[0\bar{1}1]$ &  -0.0736807 &0 &-0.0736807\\

$(111)[10\bar{1}]$ & 0.0736807 &-0.0736807& 0\\
$(1\bar{1}1)[10\bar{1}]$ &  0.0736807 &-0.0736807&0\\

$(\bar{1}11)[\bar{1}\bar{1}0]$  & 0 &-0.0736807 & -0.0736807\\
$(1\bar{1}1)[\bar{1}\bar{1}0]$ & 0 &0.0736807& 0.0736807\\

$\underline{(\bar{1}11)[101]}$ (2) & -0.0736807 &0.0736807& 0\\
$(11\bar{1})[101]$   & 0.0736807 &-0.0736807 & 0\\

$(1\bar{1}1)[0\bar{1}\bar{1}]$  & 0.0736807 &0 & 0.0736807\\  
$(11\bar{1})[0\bar{1}\bar{1}]$ & -0.0736807 &0 & -0.0736807\\\hline     
\end{tabular}
\end{center}
\end{table}

\begin{figure}
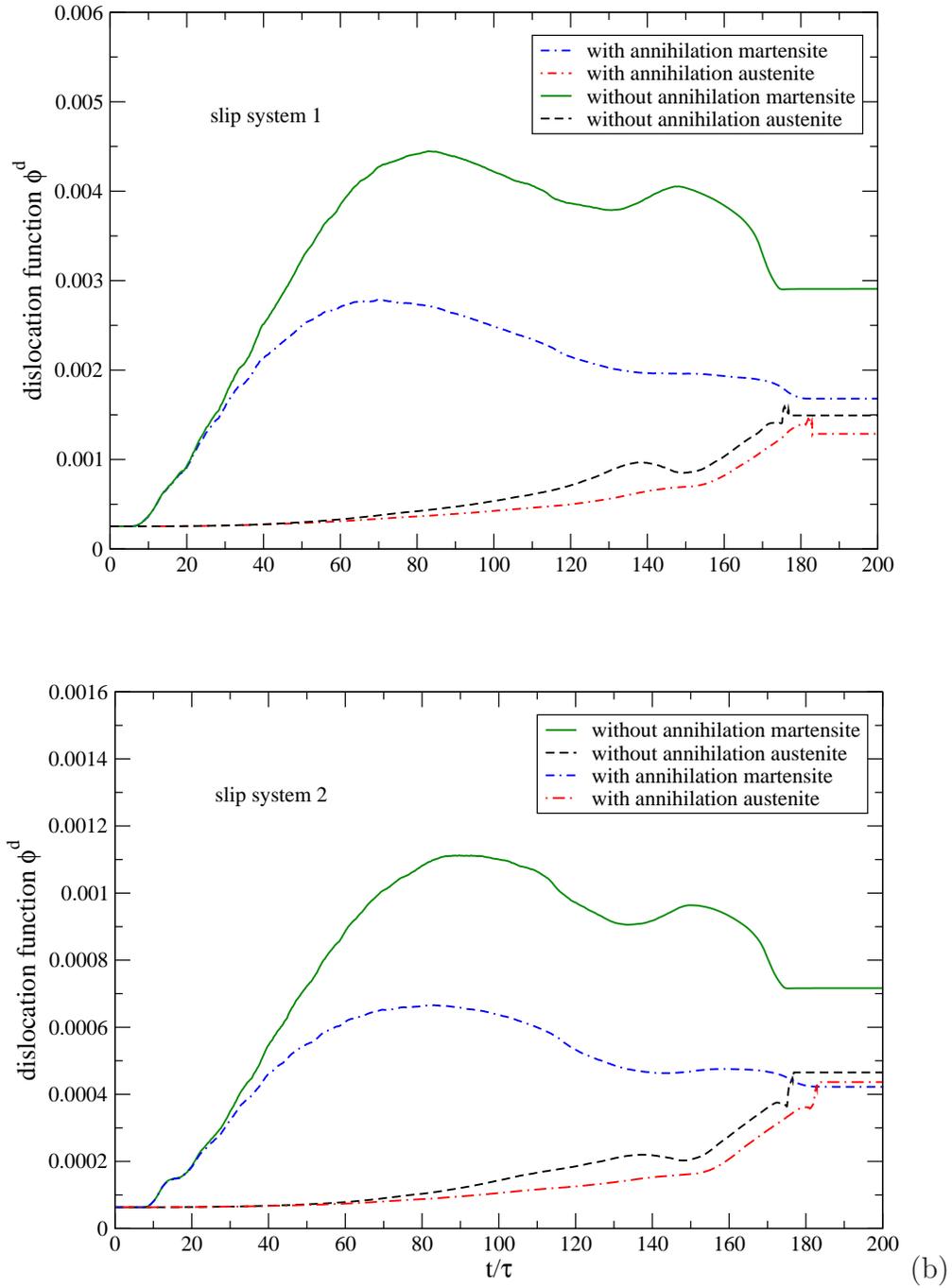

\begin{centering}
\begin{tabular}{c}
\includegraphics[scale=0.5]{Fig7a.eps}  (a)\\\\\\
\includegraphics[scale=0.5]{Fig7b.eps} (b)
\end{tabular}
\caption {\label{Fig7}
The time evolution of the dislocation density function for the slip systems 1 (a) and 2 (b).}
\end{centering}
\end{figure}

 The evolution of the mean dislocation density during the simulation is plotted in Fig.~\ref{Fig7} (a,b) for two slip systems. 
 It can be observed that the dislocation density in the martensite increases stronger than in the austenite with and without  annihilation. It is obviously that the annihilation reduces the dislocation density according to eq.~(\ref{DensEvolution}). For the slip system 1 the dislocation density is larger due to the larger summary interaction stress.

\begin{figure}
\begin{centering}
\begin{tabular}{c}
\includegraphics[scale=0.5]{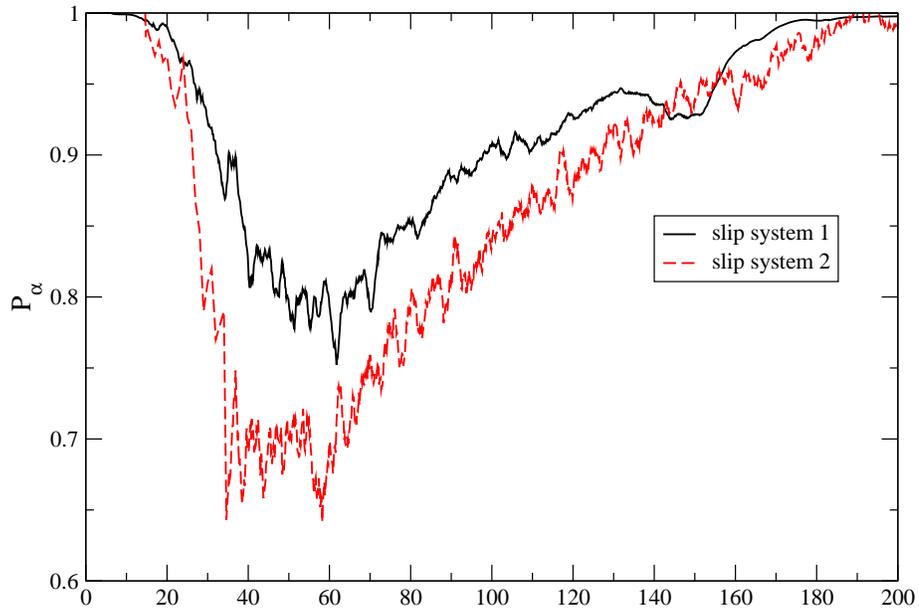}  (a)\\\\\\
\includegraphics[scale=0.5]{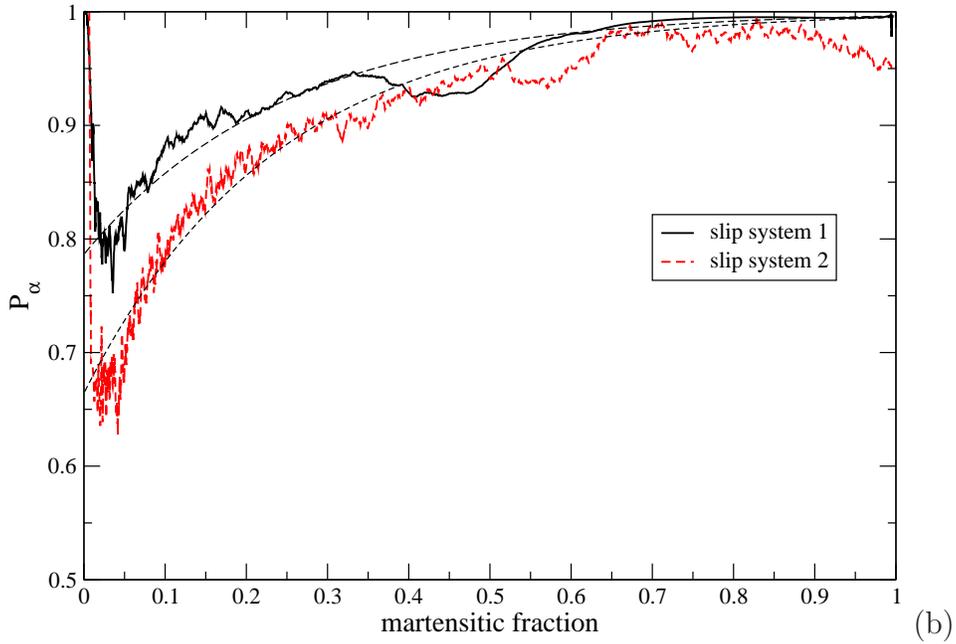} (b)
\end{tabular}
\caption {\label{Fig8}
The time evolution of the probability of the dislocation inheritance for the slip systems 1 and 2. Thin dashed lines are the approximations by eq.~(\ref{Prob_Fraction}).}
\end{centering}
\end{figure}

   The time evolution of the probability of the inheritance, $\mathcal{P}_\alpha$, is plotted in Fig.~\ref{Fig8} (a). The comparison with Fig.~\ref{Fig7} shows that the larger the dislocation density and their evolution rate, the larger the probability $\mathcal{P}_\alpha$. The test simulation with the small interface width $\xi=0.32\, l_0$ have shown that the annihilation of the dislocations at the transformation front is five times smaller. It is obviously that for a thicker interface the annihilation process is more intensive. It is important to notice that the model parameter $c_9$ is strongly connected to the interface width.  

The probability $\mathcal{P}_\alpha$ as a function of the martensite phase fraction is plotted in Fig.~\ref{Fig8} (b). 
   It strongly decreases for a short time and then increases due to the dislocation evolution in the austenite caused by the MT. At the end of the transformation the probability evolves to 1. 
In Fig.~\ref{Fig8} it can be also seen that for the slip system 1 the average probability  of the inheritance is higher than for system 2. This fact can be explained by the higher interaction stress that causes the intensive dislocation density evolution and the corresponding decrease of the contribution of the annihilation term.

 In the work \cite{Ostwald2011} a convex inheritance probability function was
 proposed, where it was required to match the inheritance probability function to  0 at $x=0$ and to 1 at $x=1$. It was also assumed that if $x$ tends to 1, the dislocation density in the austenite increases and forces to be inherited by the MT front. 
To proof the proposed convex inheritance probability function we have looked for the best approximation and found that the curves in  Fig.~\ref{Fig8} (b) can be approximated in their increasing part by a function of the martensite phase fraction, $x$
\begin{eqnarray}\label{Prob_Fraction}
\mathcal{P}_\alpha=1-(1-k_0)\exp\left(-\frac{x}{k_1}\right),
\end{eqnarray}
where $k_0$ is a minimal probability and $k_1$ is a scaling factor. For $x=1$ the probability becomes 1. 
For the simulated curves the parameters of eq.~(\ref{Prob_Fraction}) are the following:   $k_0=0.7865$, 0.6643  and $k_1=0.2462$, 0.2364 for the slip systems 1 and 2, respectively. 

\begin{figure}
\begin{centering}
\includegraphics[scale=0.5]{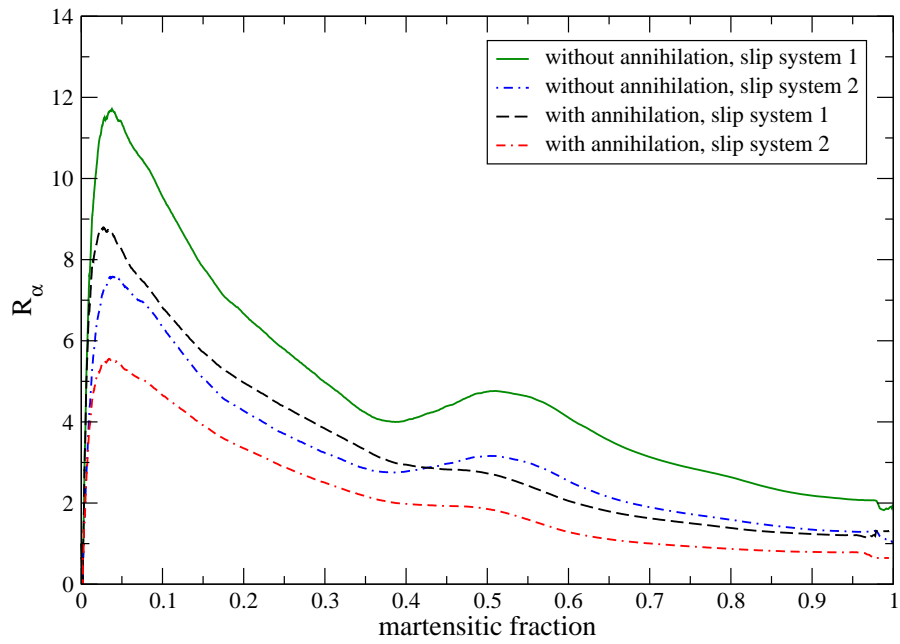}  
\caption {\label{Fig9} 
The ratio between the dislocation densities in the martensite and austenite as a function of the martensite phase fraction with and without annihilation for the slip systems 1 and 2 .}
\end{centering}
\end{figure}

The time dependency of the ratio between the dislocation densities in the martensite and austenite is plotted in Fig.~\ref{Fig9}  for two slip systems. The behavior is similar to the probability of the inheritance. With increasing dislocation density in the austenite the ratio $\mathcal{R}$ decreases. The annihilation term reduces the dislocation density in the martensite at the earlier stages and does not influence the further evolution. It is interesting that with and  without annihilation the dislocation density is higher in the martensite and the ratio  $\mathcal{R}$ is higher than 1 all the time during the MT.

\begin{figure}
\begin{centering}
\includegraphics[scale=0.5]{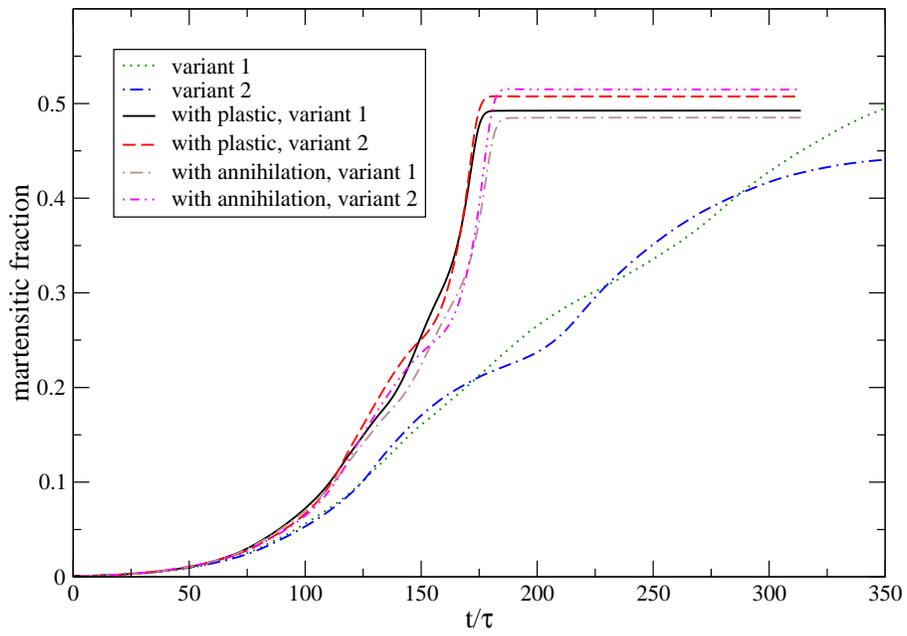} 
\caption {\label{Fig10}
Evolution of the martensite phase fraction for two variants with and without annihilation. }
\end{centering}
\end{figure}

The comparison of the kinetics of the MT with and without dislocation evolution and with and without annihilation term is shown in Fig.~\ref{Fig10}. The plastic deformation in the matrix increases the rate of the MT and the final martensite phase fraction arrives 1. The annihilation in the transformation front slightly decreases the transformation rate. 

From the simulation results it follows that at assumed parameters the contribution  of the annihilation at the transformation front is around 20-30$\%$. But in spite of this the influence of the annihilation on the MT is not essential. It is expressed in a decreasing rate of the MT and a decreasing final dislocation density almost to half.
The parameter $c_9$ can be used as a material parameter to describe the real effectiveness of the annihilation. The precise optimization of this  
parameter should be carried out by the comparison of the simulated microstructure and dislocation density field with the experimental data for example with transformation strain data measured by digital image correlation (DIC) \cite{Holzweissig12a,Holzweissig12b}.

 To show the applicability of the presented model we performed numerical simulations of a tensile test. The specimens were deformed to a true strain in $[101]$ direction. During the tensile test the relationship of the volume fraction of the stress-induced martensite $f_M$ versus true strain was studied.  The applied stress in eq.~(\ref{KineticEq}) was calculated as 
$\sigma^{\text{appl}}_{ij}=  c_{ijkl}\varepsilon^{\text{true}}_{kl}$, the undercooling was chosen $\Delta f=0$.

\begin{figure}
\begin{centering}
\includegraphics[scale=0.5]{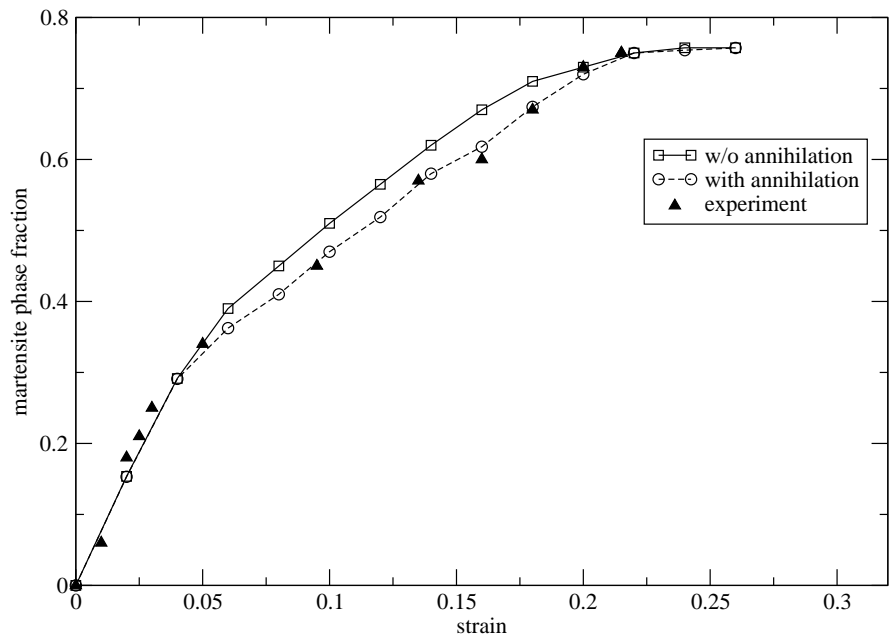} 
\caption {\label{Fig11} 
Simulated and experimental \cite{Hourman2000} dependencies of the martensite phase fraction versus the true strain. Thin solid lines are the fitting curves for the simulation results with and without the annihilation term.}
\end{centering}
\end{figure}

The final volume fractions of stress-induced martensite were found both with and without annihilation at the transformation front. 
Fig.~\ref{Fig11} shows the simulated and experimental dependencies of the final volume fraction of the stress-induced martensite as function of the true strain. The experimental results are taken from Refs. \cite{Kubler11,Hourman2000} for comparison. We found the best fit of the calculated curve to the experiment at $c_9=7.2\times 10^{-10}$ m$^{2}$.  It can be seen that the annihilation decreases the martensite fraction in the middle of the curve. A characteristic deflection at the experimental curve can be observed similar to the simulation results.  We can summarize that an appropriate choise of the parameter $c_9$ for the annihilation allows to mimic the experimental results on a tolerable level.

\section{Conclusion}
\label{Conclusion}

In the present study the interaction of the MT front with the plastic strain
in a a Fe-30$\%$Ni alloy has been investigated in detail by means of the elasto-plastic phase-field model. 

 The model takes into account the formation of accommodation dislocations in the austenitic matrix and the annihilation of the dislocations at the growing martensitic transformation front. We have suggested and tested a new  phenomenological annihilation term in the evolution equation depending on the dislocation density and the transformation rate. To study this term we define the probability of the inheritance $\mathcal{P}_\alpha$, which is equal to 1 if the annihilation term is equal to zero and is smaller than 1 otherwise. The correlation between $\mathcal{P}_\alpha$ and the martensite fraction was studied and an approximation function is found with different coefficients for the different slip systems. This new phenomenological founded dependency can then be used in the macroscopic material models for the phase transformation-plasticity interaction. In the future work the proposed model and in particular the annihilation term should be validated by the experimental microstructure and experimental 
plastic strain distribution.

\section{Acknowledgement}  We gratefully acknowledge Johannes Roedel for the fruitful discussions and scientific support.

\appendix{\textbf{Appendix A: The chemical energy function}}

The model function $f$ is defined as follows

\begin{eqnarray}
f(\vec{\eta})=\sum_{q=1}^\nu\eta_{q}^2- 2\sum_{q=1}^\nu\eta_{q}^3 +\left(\sum_{q=1}^\nu\eta_{q}^2\right)^2
\end{eqnarray}
and its derivative is equal to
\begin{eqnarray}
f'_{\eta_{p}}=2\left(\eta_{p}-3\eta_{p}^2 + 2\eta_{p}\sum_{q=1}^\nu\eta_{q}^2\right). 
\end{eqnarray}
$f$ reduces in the case of one martensitic variant to the standard double well function
\begin{eqnarray}
f=\eta^2(1-\eta)^2
\end{eqnarray}
with
\begin{eqnarray}
f'_\eta=2\eta(1-\eta)(1-2\eta).
\end{eqnarray}
For this function the numeric constant $a_1=\sqrt{2}/3$ \cite{Karma98}.

The model function $g$ is defined as
\begin{eqnarray}
g(\vec{\eta})=12\left(\frac{1}{3}\sum_{q=1}^\nu\eta_{q}^3-\frac{1}{4}\left(\sum_{q=1}^\nu\eta_{q}^2\right)^2\right)
\end{eqnarray}
with a derivative
\begin{eqnarray}
g'_{\eta_{p}}=12\left(\eta_{p}^2-\eta_{p}\sum_{q=1}^\nu\eta_{q}^2\right).
\end{eqnarray}
Further $g$ reduces in the case of the one martensitic variant to the model function
\begin{eqnarray}
g=12\left(\frac{\eta^3}{3}-\frac{\eta^4}{4}\right)
\end{eqnarray}
with
\begin{eqnarray}
g'_\eta=12\eta^2(1-\eta). 
\end{eqnarray}

It can be seen that the second and third terms in the kinetic eq.~(\ref{KineticEq}) build together the  well known  Landau-type polynomial
\begin{eqnarray}
f_L&=&Hf(\vec{\eta})-\Delta fg(\vec{\eta})\nonumber\\
&=&\frac{1}{2}a_2\sum_{q=1}^\nu\eta_{q}^2- \frac{1}{3}(3a_2+12\Delta f)\sum_{q=1}^\nu\eta_{q}^3 +\frac{1}{4}\left(2 a_2+12\Delta f\right)\left(\sum_{q=1}^\nu\eta_{q}^2\right)^2
\end{eqnarray}
where  the Landau energy parameter $a_2=2H=\frac{2\gamma}{a_1\xi}$ is related to the nucleation barrier.

\end{document}